\documentclass[smallextended]{svjour3}       
\RequirePackage{fix-cm}
\smartqed  
\usepackage{graphicx}

\usepackage{enumitem}
\usepackage{changepage} 
\usepackage[sort&compress, authoryear, comma]{natbib}
\usepackage{subfigure}
\usepackage{svg}
\usepackage[many]{tcolorbox} 
\usepackage{amsmath}

\usepackage[utf8]{inputenc} 
\usepackage[T1]{fontenc}
\usepackage{graphicx}
\usepackage{grffile}
\usepackage[bookmarksdepth=2]{hyperref} 
\usepackage[misc,geometry]{ifsym}  
\usepackage{float} 
\usepackage{ifthen} 
\usepackage{amssymb} 
\usepackage{booktabs}
\usepackage{makecell} 
\usepackage{endnotes} 
\usepackage{moresize}
\usepackage[subtle]{savetrees}
\usepackage[title]{appendix}

\usepackage{listings, xcolor}

\definecolor{verylightgray}{rgb}{0.99,0.99,0.99}

\lstdefinelanguage{Solidity}{
	keywords=[3]{block, blockhash, coinbase, def, difficulty, gaslimit, number, timestamp, msg, delegate, return,  create2, call.data, sender, sig, value, now, tx, gasprice, origin},	
	keywordstyle=[3]\color{violet}\bfseries,
	keywords=[4]{0 ,1,2,3,4,5,6,7,8,9},
	keywordstyle=[4]\color{red}\bfseries,
	identifierstyle=\color{black},
	sensitive=false,
	comment=[l]{\#},
	morecomment=[s]{/*}{*/},
	commentstyle=\color{blue}\ttfamily,
	stringstyle=\color{red}\ttfamily,
	morestring=[b]',
	morestring=[b]"
}

\lstdefinestyle{sidebyside}{
	language=Solidity,
	backgroundcolor=\color{verylightgray},
	extendedchars=true,
	basicstyle=\fontfamily{pcr}\selectfont\tiny,
	showstringspaces=false,
	showspaces=false,
	numbers=left,
	numberstyle=\tiny,
	numbersep=5pt,
	tabsize=2,
	breaklines=true,
	showtabs=false,
	captionpos=b,
	xleftmargin=3.0ex,
}

\usepackage{ragged2e} 
\usepackage[export]{adjustbox}
\usepackage{multirow} 
\usepackage{tikz}
\usepackage{url}

\usepackage{color, colortbl}
\definecolor{Gray}{gray}{0.9}
\usepackage[compatibility=false, singlelinecheck=false]{caption}
\usepackage{tabularx}

\usepackage{listings, xcolor}

\definecolor{verylightgray}{rgb}{0.99,0.99,0.99}

\lstdefinelanguage{Solidity}{
	keywords=[3]{block, blockhash, coinbase, def, difficulty, gaslimit, number, timestamp, msg, delegate, return,  create2, call.data, sender, sig, value, now, tx, gasprice, origin},	
	keywordstyle=[3]\color{violet}\bfseries,
	keywords=[4]{0 ,1,2,3,4,5,6,7,8,9},
	keywordstyle=[4]\color{red}\bfseries,
	identifierstyle=\color{black},
	sensitive=false,
	comment=[l]{\#},
	morecomment=[s]{/*}{*/},
	commentstyle=\color{blue}\ttfamily,
	stringstyle=\color{red}\ttfamily,
	morestring=[b]',
	morestring=[b]"
}

\lstdefinestyle{sidebyside}{
	language=Solidity,
	backgroundcolor=\color{verylightgray},
	extendedchars=true,
	basicstyle=\fontfamily{pcr}\selectfont\tiny,
	showstringspaces=false,
	showspaces=false,
	numbers=left,
	numberstyle=\tiny,
	numbersep=5pt,
	tabsize=2,
	breaklines=true,
	showtabs=false,
	captionpos=b,
	xleftmargin=3.0ex,
}

\makeatletter
\g@addto@macro{\UrlBreaks}{\UrlOrds}
\makeatother

\graphicspath{{Images/}}

\newcommand{\countobservations}{
    \def \countobservations{1}
}
\newcounter{observation}
\newenvironment{observation}[1]
{   \refstepcounter{observation}
    \noindent \textbf{Observation~\theobservation) 
    \textit{#1}}
}

\newboolean{showcomments}
\setboolean{showcomments}{true}

\ifthenelse{\boolean{showcomments}}
{\newcommand{\nbc}[3]{
 {\colorbox{#3}{\bfseries\sffamily\scriptsize\textcolor{white}{#1}}}
 {\textcolor{#3}{\sf\small$\blacktriangleright$\textit{#2}$\blacktriangleleft$}}
 }
}
{\newcommand{\nbc}[3]{}
 }

\newtcolorbox{mybox}[2][]{
top=0.15in,left=4pt,right=4pt,bottom=4pt,
fonttitle=\bfseries,
colbacktitle=gray,
colback=gray!5,
colframe=gray!40!black,
enhanced,
attach boxed title to top left={xshift=1.5em,yshift=-\tcboxedtitleheight/2},
boxed title style={size=small},
drop shadow={black!50!white},
title=#2,#1}

\newcommand{\countimplications}{
    \def \countimplications{1}
}
\newcounter{implication}
\newenvironment{implication}[1]
{
    \refstepcounter{implication}
    \noindent \textbf{Implication~\theimplication ) \textit{#1}}
}

\countimplications

\newcommand\PremStudy{Is the proxy pattern a relevant practice in the domain of smart contracts?}

\newcommand\RQOne{How prevalent is the proxy mechanism in the Ethereum ecosystem?}

\newcommand\RQTwo{What are the different creational patterns for deploying proxy contracts?}

\newcommand\RQThree{What are the different types and properties of proxy contracts?}

\begin{document}

\title{A Large-Scale Exploratory Study on the Proxy Pattern in Ethereum
}

\author{Amir~M.~Ebrahimi \and Bram~Adams \and Gustavo~A.~Oliva \and Ahmed~E.~Hassan
}


\institute{Amir~M.~Ebrahimi, Gustavo~A.~Oliva, Ahmed~E.~Hassan \at Software Analysis and Intelligence Lab (SAIL), School of Computing \\
Queen’s University, Kingston, Ontario, Canada \\
\email{\{amir.ebrahimi,gustavo,ahmed\}@cs.queensu.ca}           
           \and
           Bram~Adams \at Lab on Maintenance, Construction and Intelligence of Software (MCIS), School of Computing \\
Queen’s University, Kingston, Ontario, Canada \\
\email{\{bram.adams\}@cs.queensu.ca}
}

\date{Received: date / Accepted: date}

\maketitle
    \begin{abstract}
    The proxy pattern is a well-known design pattern with numerous use cases in several sectors of the software industry (e.g., network applications, microservices, and IoT). As such, the use of the proxy pattern is also a common approach in the development of complex decentralized applications (DApps) on the Ethereum blockchain. A contract that implements the proxy pattern (proxy contract) acts as a layer between the clients and the target contract, enabling greater flexibility (e.g., data validation checks) and upgradeability (e.g., online smart contract replacement with zero downtime) in DApp development. Despite the importance of proxy contracts, little is known about (i) how their prevalence changed over time, (ii) the ways in which developers integrate proxies in the design of DApps, and (iii) what proxy types are being most commonly leveraged by developers. In this paper, we present a large-scale exploratory study on the use of the proxy pattern in Ethereum. We analyze a dataset of all Ethereum smart contracts as of Sep. 2022 containing 50M smart contracts and 1.6B transactions, and apply both quantitative and qualitative methods in order to (i) determine the prevalence of proxy contracts, (ii) understand the ways they are deployed and integrated into applications, and (iii) uncover the prevalence of different types of proxy contracts. Our findings reveal that 14.2\% of all deployed smart contracts are proxy contracts. We show that proxy contracts are being more actively used than non-proxy contracts. Also, the usage of proxy contracts in various contexts, transactions involving proxy contracts, and adoption of proxy contracts by users have shown an upward trend over time, peaking at the end of our study period. They are either deployed through off-chain scripts or on-chain factory contracts, with the former and latter being employed in 39.1\% and 60.9\% of identified usage contexts in turn. We found that while the majority (67.8\%) of proxies act as an interceptor, 32.2\% enables upgradeability. Proxy contracts are typically (79\%) implemented based on known reference implementations with 29.4\% being of type ERC-1167, a class of proxies that aims to cheaply reuse and clone contracts' functionality. Our evaluation shows that our proposed behavioral proxy detection method has a precision and recall of 100\% in detecting active proxies. Finally, we derive a set of practical recommendations for developers and introduce open research questions to guide future research on the topic.

\end{abstract}

    \section{Introduction}
\label{sec:introduction}

A proxy is fundamentally an entity that acts as an intermediary between two objects~\citep{gof_book95}. It serves as a representative for a target object—whether it is a network connection, a large memory object, a file, or other resource-intensive entities—without altering its interface. Hence, for a client, interacting with a proxy feels akin to using the actual object~\citep{gof_book95}.

Proxies play a crucial role in software design, evidenced by their inclusion as one of the twenty-three standard design pattern~\citep{gof_book95}. They find applications across various software domains, including the internet, microservices~\citep{Richardson18}, IoT~\citep{Bloom18, Ngaogate19}, and blockchain~\citep{Wohrer18}. Proxies enhance modularity and encapsulation while offering functionalities like caching for resource-intensive operations, precondition checks, and access control to the real object's operations.

Previous studies have emphasized the significance of the proxy pattern in programmable blockchains, especially Ethereum~\citep{Wohrer18, Rajasekar20, Xu21}. Ethereum, a leading blockchain platform, enables the deployment of smart contracts, typically written in Solidity. In the remainder of this paper, the term smart contract will be used to refer to deployed contracts. Once these contracts are deployed, they become immutable, meaning they cannot be altered post-deployment. However, like all software, smart contracts need maintenance. Proxies provide a method for seamless upgrades to these contracts. When a logic contract (a.k.a., the actual serving object) requires an upgrade, a fresh version is deployed, and the proxy contract is adjusted to point to this new version, ensuring continuous interaction without any service disruption. This redirection is achieved by updating the proxy's reference to the new logic contract, allowing clients to maintain interaction without altering their existing setups.

While the benefits of proxies are well-established in traditional applications, their merits in blockchain applications remain under exploration. Existing smart contract literature indicates that proxies can reduce deployment costs and facilitate maintenance~\citep{Wohrer18, worley2019, Rajasekar20, Xu21, Kannengieser22, eip_897}. Nevertheless, it remains unclear i) how widely proxies are used in blockchain applications, ii) how developers effectively integrate proxies into existing applications, and iii) what types of proxies are most commonly used in practice.

Furthermore, tracking proxies is crucial for the security of many blockchain applications~\citep{eip1967}. Since the upgrade process involves modifying the code, it must be done in a secure and controlled manner to prevent any unintended consequences. This activity enables developers, auditors, and clients to monitor smart contract evolution, ensuring changes are conducted securely and transparently, which is vital for critical or high-value contracts. Currently, however, there is a lack of techniques that can effectively detect proxies at \textit{scale} and in a \textit{timely} manner. 

The aforementioned challenges underscore the imperative for in-depth exploration in this domain. Thus, we set out to conduct a large-scale exploratory study to dissect the proxy pattern in smart contracts. Our study encompasses the entire Ethereum lifespan as of Sep-01-2022 with 50,845,833 deployed smart contracts and 1,695,517,186 performed transactions. In particular, we designed an approach that effectively and efficiently mines all active proxy contracts based on their behavior, after which we addressed the following research questions:

\smallskip \noindent \textbf{Motivation-study. \PremStudy}

\smallskip \noindent \textbf{RQ1. \RQOne} 

\smallskip \noindent \textbf{RQ2. \RQTwo} 

\smallskip \noindent \textbf{RQ3. \RQThree} 

\smallskip \noindent \textbf{Contributions}. To the best of our knowledge, this is the first in-depth, large-scale exploratory study of the proxy pattern in smart contracts. In highlighting our contributions, we emphasize that the value of our work lies not just in its findings and implications, but also in its adaptable methodology and the new research questions it introduces for future exploration. More specifically, our key contributions are as follows: \textbf{(i)} proposing an efficient and accurate method for detecting active proxies, \textbf{(ii)} illustrating the prevalence of the proxy pattern and their role in the blockchain context from various perspectives, \textbf{(iii)} mining different creational patterns for deploying proxies and shedding light on the main practices, \textbf{(iv)} analyzing the prevalence of different proxy types in terms of their purpose and implementation types, \textbf{(v)} publishing four ground truth datasets for evaluating proxy detection, and proxy type detection methods, \textbf{(vi)} analyzing the practical application of the proxy pattern in the real-world decentralized applications (DApps), and \textbf{(vii)} a detailed discussion on the pros and cons of the proxy pattern and different proxy deployment styles, and the ensuing research challenges and practical implications.

\smallskip \noindent \textbf{Paper structure}. Section~\ref{sec:background} discusses key relevant terminologies and concepts used across the paper. Section~\ref{sec:data-collection} discusses our data collection. Section~\ref{Sec:prem} describes a preliminary analysis to further enforce the study of the proxy pattern in Ethereum. Section~\ref{sec:proxy-detection-and-eval} introduces our proxy detection method and its evaluation. Sections~\ref{sec:rq1}, \ref{sec:rq2}, and \ref{sec:rq3} describe our motivation, approach, and findings for each of our three research questions, respectively. Section~\ref{sec:discussion} compares and describes the benefits and drawbacks of adopting the proxy pattern and its two deployment styles.  Next, the implications of our findings as well as opportunities for future research are described. Section~\ref{sec:threats} lists the threats to the validity of our study. Section~\ref{sec:related-work} surveys the relevant literature. Finally, Section~\ref{sec:conclusion} concludes our study.

    \section{Background}
\label{sec:background}




\noindent \textbf{Blockchain.} A blockchain is a distributed ledger managed by nodes in a peer-to-peer network. Notable platforms include Bitcoin, Ethereum, EOS, POA, Nxt, and Hyperledger Fabric. Ethereum stands out as a leading programmable blockchain, facilitating the hosting and execution of smart contracts through its Ethereum Virtual Machine (EVM).

\smallskip \noindent \textbf{Smart contract.} A smart contract is a program executed on a blockchain, commonly written in Solidity, a language with syntax akin to JavaScript. Once compiled, the bytecode is deployed to platforms like Ethereum, which typically do not store the contract's source code. In this study, the term "smart contract" refers to this bytecode. On the Ethereum blockchain, each deployed contract is assigned a unique address for identification. A \textit{verified} smart contract has its source code publicly accessible on Etherscan\footnote{\url{https://etherscan.io/contractsVerified}} for review.



\smallskip \noindent \textbf{Decentralized Application (DApp).} A decentralized application operates on programmable blockchains like Ethereum, which offers a secure and transparent platform. These DApps utilize smart contracts to manage transactions, ensuring their back-end code remains open-source and not controlled by any single entity. The Ethereum network verifies and processes all transactions, providing resistance to censorship, fraud, and downtime. DApps have diverse applications such as decentralized finance (DeFi), gaming, gambling, etc.


\smallskip \noindent \textbf{Account types.} The Ethereum platform supports two types of accounts, namely externally owned accounts (EOA) and contract accounts (CA). An EOA contains the following fields: an address (40-digit hexadecimal ID), a transaction counter, and the ETH balance (ETH is the official Ethereum cryptocurrency). A contract account, in turn, holds the bytecode of a smart contract in addition to the previously mentioned fields.

\smallskip \noindent \textbf{Transactions and contract deployment.} In Ethereum, transactions facilitate interactions and are exclusively initiated by EOAs. EOAs transfer cryptocurrency (Ether) to other EOAs or send transactions to deploy or invoke functions in contracts. Notably, contracts can also engage with other contracts, a feature facilitated by internal transactions, hereafter referred to as traces. A trace encapsulates details of an operation that took place during transaction execution, including its type, involved addresses, execution state, calldata, output data, transferred Ether amount, and gas consumption. 

\begin{figure}[!t]
\centering
\includegraphics[width=1.0\columnwidth]{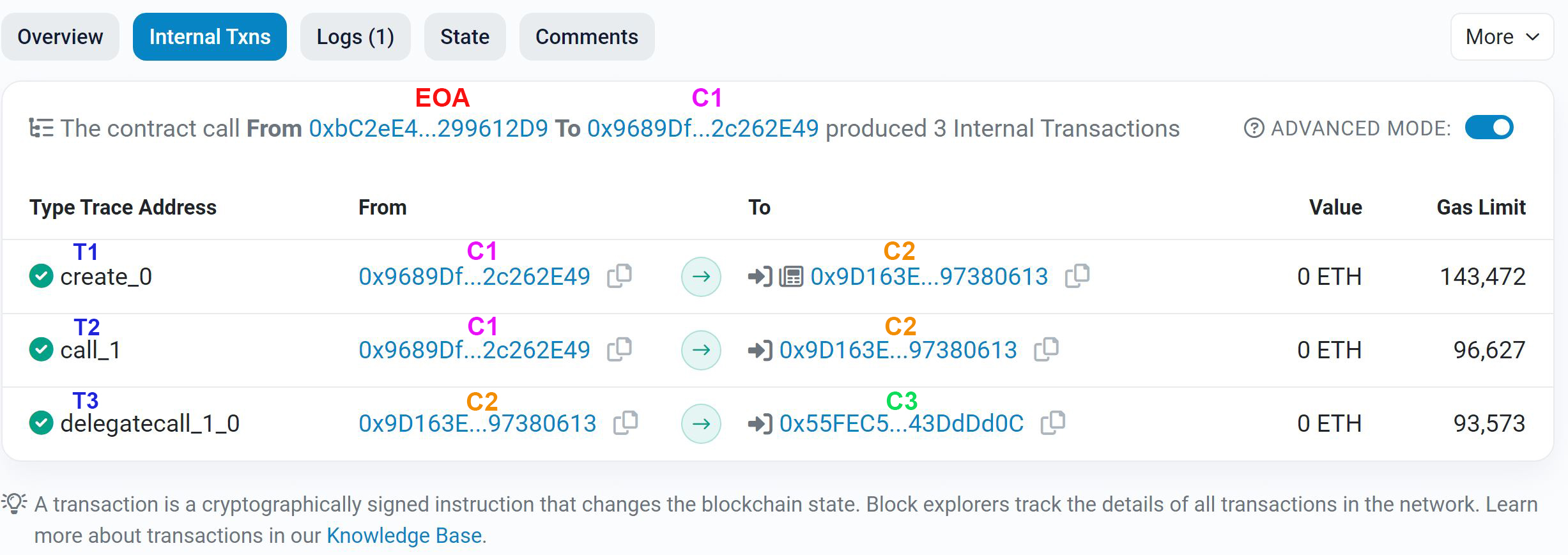} 
\caption{The trace table of an example transaction \textit{T}.}
\label{fig:trace-table} 
\end{figure}

Transactions are the means through which one interacts with Ethereum. Transactions are always initiated by an EOA. EOAs can send transactions to other EOAs to transfer cryptocurrency (Ether). However, since Ethereum is a programmable blockchain, EOAs can send transactions to (i) deploy contracts and (ii) interact with contracts (i.e., invoke functions defined in the API of deployed contracts). Interestingly, contracts themselves can also deploy and interact with other contracts. These functionalities are enabled by a special mechanism known as internal transactions. We refer to an internal transaction as a trace, each representing an operation that took place during transaction execution. A trace contains information about various aspects of the operation, such as its type, the caller and callee addresses (i.e., \texttt{From} and \texttt{To} addresses, respectively, in Figure \ref{fig:trace-table}), execution state, calldata (i.e., the callee's function selector and parameters), output data, the amount of Ether transferred, and the gas used. 

\begin{figure}[!t]
 \centering
 \includegraphics[width=0.30\columnwidth]{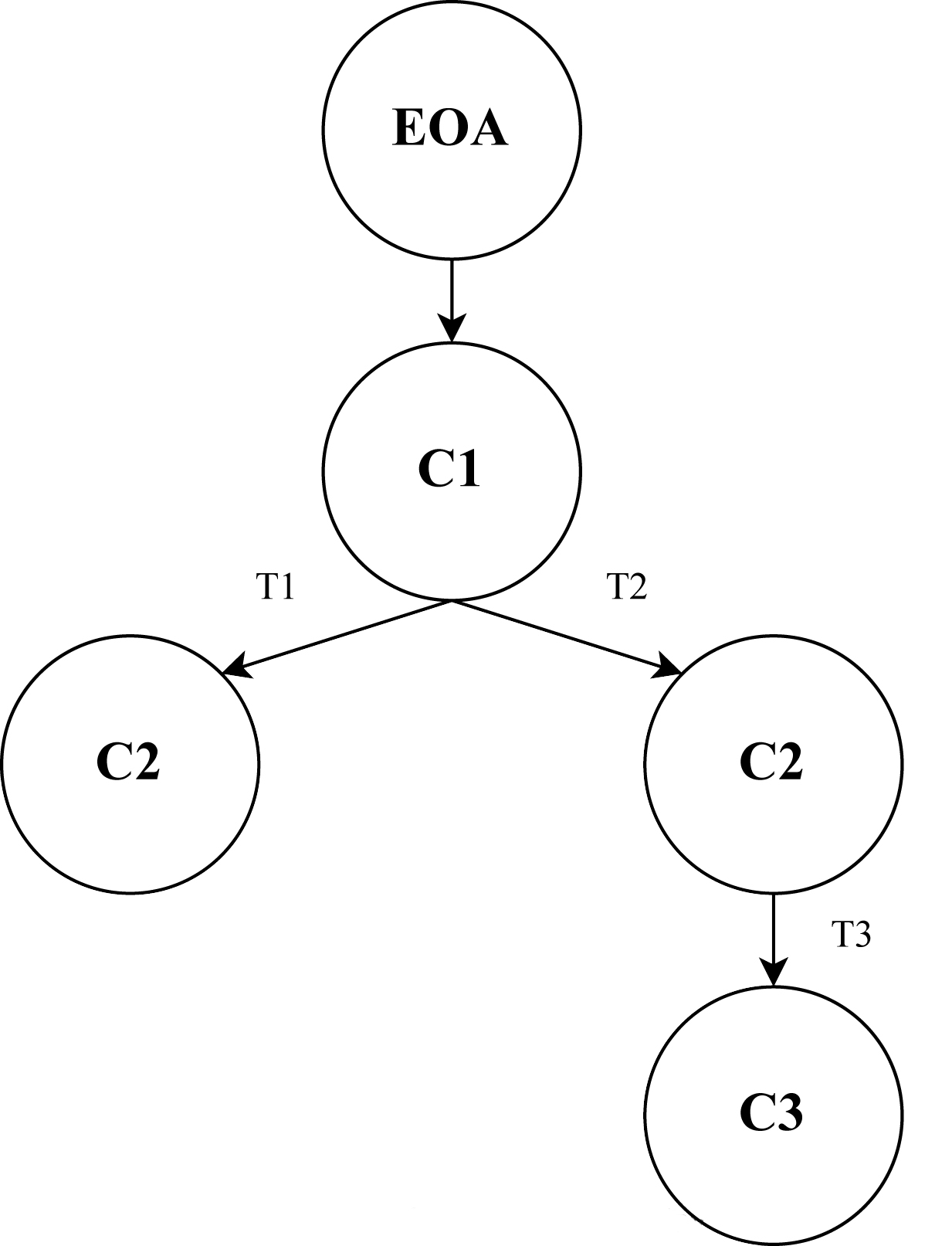}
 \caption{The call graph for an example transaction \textit{T}.}
 \label{fig:trace-call-graph}
\end{figure}

Figure \ref{fig:trace-table} depicts a trace table for an example transaction\footnote{\url{https://etherscan.io/tx/0x650e7876a7742194b14544bc7ed2e9b8fb9a63861e24f83e9bd3ee4d92673fa5/advanced\#internal}} \textit{T}. For the purpose of readability, we labeled the EOA, involved contracts and each trace with appropriate abbreviations denoted by different colors in Figure \ref{fig:trace-table}. As shown, as a result of a call from the EOA to the contract C1, three traces, namely \textit{T1}, \textit{T2}, and \textit{T3}, were executed. In particular, \textit{C1} creates another contract \textit{C2} (i.e., \textit{T1}). Then, \textit{C1} makes a call to \textit{C2} (i.e., \textit{T2}), which resulted in another child trace (i.e., \textit{T3}), where \textit{C2} delegated a call to \textit{C3}. Thus, a trace can have child sub-traces. Figure \ref{fig:trace-call-graph} shows the call graph for transaction \textit{T}, denoting the precedence of traces against each other. The direction of the arrows indicates the parent-child relationship between traces. For instance, trace \textit{T3} is a child of trace \textit{T2}; thus, \textit{T2} precedes \textit{T3}.

Finally, when a contract is deployed by an EOA, it often means that an entity (commonly a developer) deployed the contract to Ethereum. When a contract \textit{C2} is deployed by another contract \textit{C1}, it means that \textit{C1} deployed \textit{C2} at runtime as a result of a transaction \textit{T} sent to \textit{C1}. The definition of \textit{C2} can be either given to \textit{C1} (e.g., as part of the calldata field contained in \textit{T}) or created dynamically by \textit{C1}.

\smallskip \noindent \textbf{Transaction payment (gas system).} To successfully conduct a transaction on Ethereum, an EOA must pay a transaction fee. This fee is a function of (i) the number and type of smart contract instructions that are executed during runtime (a.k.a., the \textit{gas used}) and (ii) a financial incentive for miners to process the transaction (a.k.a., gas price). More details about the gas system can be found in the work of~\citet{Oliva21}.

\smallskip \noindent \textbf{Cross-contract method calls.} The Solidity language supports four types of cross-contract method calls, namely \textit{call}, \textit{callcode}, \textit{staticcall}, and \textit{delegatecall}. To explain the differences among these types, take Alice, Bob, and Charlie as three contracts. When Alice does a \textit{call} to Bob, the code runs in the context of Bob; thus, Bob's storage is prone to change. The \textit{staticcall} was a security update that allows calling another contract while prohibiting any state changes during the call (and its sub-calls). When Alice calls Bob and Bob does \textit{callcode} to Charlie, the code runs in the context of Bob, which means Charlie can change Bob's storage. Here, Charlie sees Bob as the message sender. The \textit{delegatecall} is often seen as a bug fix for \textit{callcode}, since the latter does not preserve the message sender. This means that Charlie sees Alice instead of Bob as the message sender if we replace \textit{callcode} with \textit{delegatecall} in the previous scenario. 

\smallskip \noindent \textbf{Ethereum Request for Comments (ERCs).} 
An ERC (Ethereum Request for Comment) is a proposal in the Ethereum community that describes application-level standards and conventions for the Ethereum platform.

\begin{figure}[!t]
 \centering
 \includegraphics[width=0.85\columnwidth]{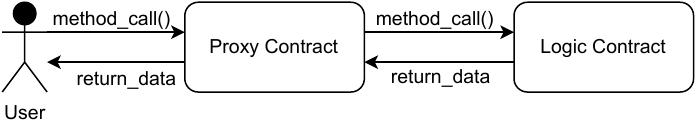}
 \caption{The basic proxy contract design.}
  \label{fig:back-01}
\end{figure}

\smallskip \noindent \textbf{Proxy contract.} A proxy contract, commonly known as a dispatcher, embodies the proxy pattern, as shown in Figure \ref{fig:back-01}. An external entity (either an EOA or a CA) initiates a function call to the proxy. The proxy, in turn, utilizes a \textit{delegatecall} to redirect this call to the logic contract, which houses the operational code. This logic contract's reference is maintained within the proxy's storage. The outcome is relayed back to the caller through the proxy. The main point in the proxy design is that, from the external entity's perspective, talking to the proxy must resemble talking to the actual logic contract. Therefore, the proxy and its logic must follow a similar interface. This mandates that function selectors\footnote{Defined as the initial four bytes of the keccak256 hash of a function's signature.} in both contracts align. Thus, two distinguishing features of proxy contracts emerge: \textbf{i)} the use of \textit{delegatecall} for interactions and \textbf{ii)} a shared interface with its logic contract. These features guide our proxy contract identification in Section \ref{sec:proxy-detection}. Proxies can function as either a \textit{forwarder} or an \textit{upgradeability} proxy. The former retains a fixed reference to its logic contract post-deployment, while the latter permits reference updates to new logic contracts without downtime.

\smallskip \noindent \textbf{Proxy pattern standards in Ethereum.} The Ethereum community has been proactive in devising standards for proxy contract implementation and management. ERC-897 introduced an interface that standardizes operations across diverse proxy types, highlighting two primary kinds: the forwarder proxy and the upgradeability proxy~\citep{eip_897}. Concurrently, ERC-1967 responded to the need for secure proxy and logic contract address tracking, proposing standard storage slots for both the logic contract's address and the proxy's administration. This standardization effort aimed to boost tool compatibility across the platform~\citep{eip1967}.

Complementing these initiatives, ERC-1167 proposed the "Minimal Proxy Contract" to enable the cost-effective cloning of contract functionalities~\citep{eip_1167}, while ERC-1822's Universal Upgradeability Proxy Standard (UUPS) streamlined the developer experience in managing proxy and logic contracts, vesting the upgrade initiation in the logic contract\citep{eip_1822}. The Diamonds standard further introduced a flexible structure, allowing for multiple facets and the addition or replacement of multiple functions in a single transaction~\citep{EIP-2535}. Additionally, novel approaches like the Beacon Proxy Pattern~\citep{beacon_proxy} and OpenZeppelin's Transparent Proxy Pattern emerged~\citep{openzeppelin_lab}, addressing collective upgrade challenges and selector clashing issues, respectively. These advancements underscore the Ethereum community's recognition of the significance of proxy contracts and their critical role in ensuring flexibility, upgradeability, and security within decentralized applications.

\smallskip \noindent \textbf{Alternative smart contract upgradeability techniques}. In the realm of smart contract upgradeability, various strategies exist. The Data Separation method decouples data and logic, maintaining data in a consistent contract and updating only the logic when needed, exemplified by Eternal storage \citep{smartcontract_upgrades}. The Create2 mechanism leverages the \texttt{CREATE2} method to pre-determine contract addresses and allows re-deployment at the same address under certain conditions, ensuring continuity \cite{Feist19}. Data migration involves deploying a new contract version and transferring existing data, demanding meticulous management to prevent data inconsistencies \citep{qasse2023}. The Strategy pattern, borrowed from object-oriented programming, offers a modular approach by encapsulating algorithms that can be swapped without altering the core contract \citep{smartcontract_upgrades}. Lastly, the \texttt{SelfDestruct} mechanism in Solidity enables contracts to be destroyed and subsequently replaced, though it poses challenges in data migration and irreversible actions, making it a less favored approach \citep{Chen20-2}. Among these, the proxy pattern (Figure \ref{fig:back-01}) stands out as the most popular due to its versatility and efficiency in maintaining contracts \citep{Salehi22, openzeppelin_proxy}.

\smallskip \noindent \textbf{Proxy Storage Slots.} To ensure the integrity of data between a proxy and its implementation contract, it is crucial to avoid storage conflicts. Potential issues arise when both the proxy and implementation contracts, unknowingly, use the same storage slot for different purposes. Several standards have been proposed to standardize the storage designs of these contracts, reducing storage collision risks. Key among these are the ERC-1967 Proxy Storage Slot \citep{eip1967}, which prescribes specific locations for proxies to store addresses; the Inherited Storage Proxy, emphasizing the synchronization of storage structures between proxy and implementation contracts \citep{openzeppelin_2020_proxy_pattern}; the Eternal Storage Proxy, which establishes a consistent storage schema across contract versions\citep{openzeppelin_2020_proxy_pattern}; and the Unstructured Storage Proxy, which uses a constant variable hashing approach to define storage slots, ensuring that successor versions remain compatible with their predecessors \citep{openzeppelin_2020_proxy_pattern}.

    \section{Data Collection}
\label{sec:data-collection}

Our study requires data about Ethereum smart contracts and their associated transactional activities. We collected this data using Google BigQuery, a fully managed, cloud-native data warehouse offered by the Google Cloud Platform. Google BigQuery is maintained by Google and is a paid service, although they offer a free tier with limited capabilities. We used a free tier account to collect our datasets. More specifically, we extracted data from the crypto\textunderscore ethereum dataset. This dataset, which is part of the BigQuery Public Datasets program, records information about transactions, contracts, blocks, events, tokens, and token holders. Users can extract data from this dataset using SQL-like queries, which can be executed directly within the BigQuery Console or by making API calls using the BigQuery API. 



To conduct our empirical analyses, we extracted data from the \texttt{contracts}\footnote{bigquery-public-data.crypto\textunderscore ethereum.contracts} and \texttt{traces}\footnote{bigquery-public-data.crypto\textunderscore ethereum.traces} tables. While the former contains information regarding the deployed smart contracts (e.g., address, bytecode, creation timestamp, etc.), the latter records detailed information about smart contract transactions (e.g., transaction hash, input, output, etc.).

Our empirical analyses were performed on a snapshot of the aforementioned tables, which comprised data from Aug-07-2015 to Sep-01-2022 (i.e., the time when we started our experiments). Specifically, the \texttt{contracts} table contains records of 50,845,833 distinct smart contracts, whereas the \texttt{traces} table contains information about 1,695,517,186 distinct transactions as well as 5,503,071,306 traces as of the mentioned data collection date.

Throughout our paper (Sections~\ref{Sec:prem},~\ref{sec:evaluation}, and~\ref{sec:rq3}), we applied Cochran's sample size formula to accurately determine the needed sample size for representative results~\citep{samplingmethod2021}. The inputs to this formula are confidence level and confidence interval (a.k.a., margin of error). In empirical software engineering research, a sample size is typically deemed representative (enough) when its associated confidence level and interval are 95\% and 5 respectively~\citep{pacheco22, lin23, tagra2022revisiting}. 
    \section{Motivation study: \PremStudy} 
\label{Sec:prem}

\noindent \textbf{Motivation.} To motivate our exploratory study of the proxy pattern, we first perform a preliminary study to examine if there is any sign of proxies in smart contracts. 

\begin{figure}[!t]
\begin{minipage}[t]{0.49\linewidth} 
 \includegraphics[width=\linewidth]{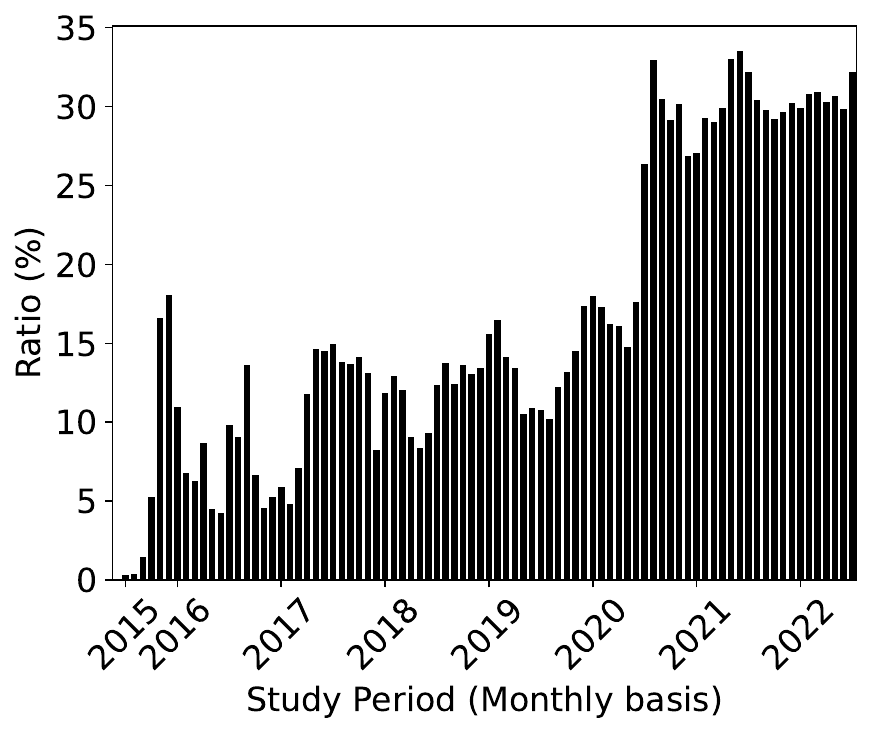}
 \caption{The monthly ratio of multi-contract transactions.}
  \label{fig:prem-01}
\end{minipage}\hfill 
\begin{minipage}[t]{0.49\linewidth}
 \includegraphics[width=\linewidth]{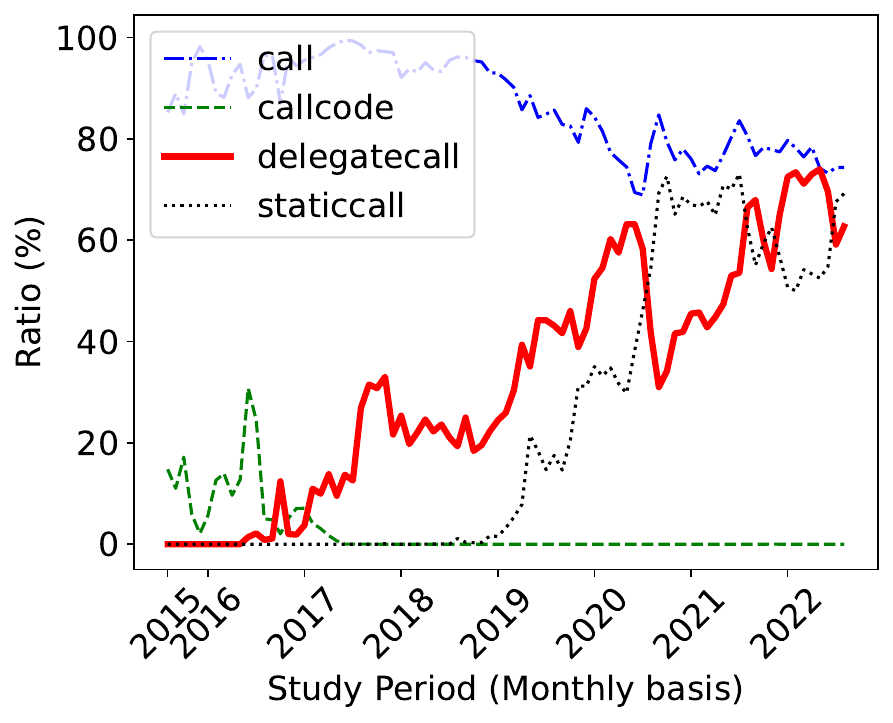}
 \caption{The monthly ratio of multi-contract transactions with different types of calls.}
  \label{fig:prem-03}
\end{minipage}
\end{figure}

\smallskip \noindent \textbf{Approach.} Given the importance of proxies in enhancing software system modularity and encapsulation~\citep{Shapiro86}, we aimed to identify signs of modularity by studying contract transactions that reflect the distribution of roles and responsibilities among various contracts. To achieve this, we queried the ``traces'' table. Specifically, we computed the monthly ratio of transactions that featured at least one cross-contract call between two different smart contracts (Figure \ref{fig:prem-01}). These transactions are referred to as ``multi-contract transactions''. It is worth noting that a multi-contract transaction can involve an arbitrary number of calls, all utilizing Solidity cross-contract call types (Section \ref{sec:background}).
Since proxy contracts typically employ the \texttt{delegatecall} type of calls in their design~\citep{openzeppelin_2020_proxy_pattern}, we assessed the monthly ratio of multi-contract transactions using each of the four cross-contract call types. This analysis helped us gauge the trend in the adoption of \texttt{delegatecall}. To verify whether a proxy contract was genuinely involved, we leveraged Etherscan's Proxy Verification\footnote{\url{https://etherscan.io/proxyContractChecker}}. We automatically analyzed a statistically representative sample of multi-contract transactions containing at least one \texttt{delegatecall} operation. In cases where a multi-contract transaction featured multiple \texttt{delegatecall} operations, we randomly selected one. Our sample size determination was based on a 95\% confidence level and a confidence interval of 5.

\smallskip \noindent \textbf{Findings.} \observation{Multi-contract transactions are becoming more widespread, with almost one-third of transactions involving at least two contracts in 2022.}Figure~\ref{fig:prem-01} illustrates a notable rise in the ratio of multi-contract transactions. Initially, in 2015, multi-contract transactions were virtually non-existent, but a dramatic surge was seen by the end of the year, with December hitting a notable 16.54\% ratio. Throughout 2016 to 2019, the ratio meandered between 4\% to 17\%, reflecting some fluctuations but a general acceptance and steady usage of multi-contract transactions. However, 2020 marked a significant year with an average ratio soaring past 25\%, and by August reaching an astonishing 32.91\%. This uptrend persisted into 2021, maintaining an average above 30\%. As of the recent data in 2022, the trend continues to hover around 30\%, suggesting a consistent adoption rate and reliance on multi-contract transactions in the Ethereum ecosystem. This long-term observation underscores the growing complexity and interconnectedness of contracts on the platform over the years.

\smallskip \observation{The prevalence of multi-contract transactions with a delegatecall operation (i.e., the primary signature of proxy contracts) is exhibiting a notable upward trajectory.}As applications grow in complexity and modularity, developers may find it impractical to hardcode all communications (i.e., regular calls) among contracts. Solidity's management of contracts' states during interactions further underscores this challenge. For instance, while regular calls typically alter the state of the callee contract, some applications necessitate changes to the caller's state or none of the contracts' states in a call chain. This necessitated the introduction of different cross-contract method calls (Section \ref{sec:background}). Figure~\ref{fig:prem-03} illustrates the monthly ratio of multi-contract transactions using one of the four call types. More specifically, from Aug. 2015 to Sep. 2022, a shift in the multi-contract transaction call types was observed. While the \texttt{call} type began with a strong preference, hovering above 85\% in 2015 and persisting through 2017, its use diminished by 2022, stabilizing around the mid-70s. In contrast, the \texttt{delegatecall} type witnessed a substantial rise, remaining negligible until May 2016 but surging consistently thereafter, surpassing 70\% by the end of the study period. The \texttt{staticcall} was initially introduced in Oct. 2017 and, remained dormant until mid-2018 but later gained traction rapidly and remained prevalent until Sep. 022. On the other hand, the \texttt{callcode} type saw an initial spike in 2015, but its utilization dwindled significantly, becoming almost negligible from 2017 onwards, indicating a decreasing inclination towards this call type.

\smallskip \observation{98.4\% of multi-contract transactions with delegatecall operations involve a proxy contract.} The increasing use of \texttt{delegatecall} operations prompted our investigation. From a dataset of 190,493,052 multi-contract transactions containing at least one \texttt{delegatecall}, we drew a statistically representative random sample. Using a confidence level and interval of 95\% of 5 respectively (sampling parameters), we obtained 385 \texttt{delegatecall} operations. For each \texttt{delegatecall}, we identified the contract that initiated it, referred to as the ``caller'' (e.g., in the third trace of Figure \ref{fig:trace-table}, \textit{C2} is the caller). We categorized the caller as either ``proxy'' or ``non-proxy'' using Etherscan's Proxy Verification service. We found that proxy contracts are predominant (98.4\%), with the remaining 1.6\% involving no proxy contracts. This high prevalence of proxies motivates us to conduct a more comprehensive study, exploring their prevalence, creation patterns, and types in further detail (Sections~\ref{sec:rq1}, \ref{sec:rq2} and \ref{sec:rq3}).

\begin{footnotesize}
    \begin{mybox}{Summary}
        \textbf{Motivational study: \PremStudy}
        \tcblower An analysis of smart contracts at the transaction level has revealed an increasing trend of up to 33\% monthly in the ratio of transactions with multiple contracts. A closer look at the type of calls in these multi-contract transactions indicates an increasing relevance of the \texttt{delegatecall} type of call. Most importantly, 98.4\% of multi-contract transactions with \texttt{delegatecalls} involve a proxy contract.
    \end{mybox}
\end{footnotesize}
    \section{Proxy Detection Approach \& Evaluation}
\label{sec:proxy-detection-and-eval}

In this section, we introduce our proxy detection approach (Section~\ref{sec:proxy-detection}) and subsequently evaluate its performance (Section~\ref{sec:evaluation}).

\subsection{Detection approach}
\label{sec:proxy-detection}

In our preliminary study (Section \ref{Sec:prem}), we initially used Etherscan's Proxy Verification service to identify proxies. However, as we scaled up our analyses for the remainder of this paper, we found this service to be impractical. Consequently, we adopted a more efficient behavioral approach.

Specifically, considering the operational characteristics of proxy contracts (Section \ref{sec:background}), we leveraged \textit{delegatecall} traces to identify proxies based on their runtime behavior. We termed the contract initiating a \textit{delegatecall} trace as the ``candidate proxy''. For instance, in trace \textit{T3} from Figure \ref{fig:trace-table}, \textit{C2} serves as the candidate proxy. To confirm whether a candidate proxy is indeed a proxy, we analyzed the preceding trace (\textit{T2} in this example) to check if its function selector (the four bytes of the trace's calldata) matched that of the \textit{delegatecall} trace. A match indicated that the proxy was calling the same function as its caller, thus confirming it as a proxy; otherwise, it was discarded.

We operationalized this procedure into a BigQuery SQL query and applied it to assess the behavior of 50,845,833 smart contracts deployed on Ethereum as of Sep. 2022 via Google BigQuery console. It is important to note that our proposed behavioral approach relies on a contract's transactional activities (runtime behavior) to determine its proxy status. Consequently, it may not identify a proxy that has never been utilized post-deployment (Section \ref{sec:evaluation}). However, this approach offers the advantage of not relying on a contract's source code or bytecode representation, making it more versatile and applicable to a broader range of scenarios.

\subsection{Evaluation}
\label{sec:evaluation} 
To evaluate our proxy detection approach, we followed these key steps. Initially, we created a ground truth dataset by selecting a statistically representative sample from a pool of 50 million deployed contracts as of September 2022. The sample size, determined with a 95\% confidence level and a confidence interval of 5, amounted to n = 385. 

Second, we conducted an analysis of the decompiled \textit{bytecode} of the sampled contracts to categorize them as either ``proxy'' or ``other''. To achieve this, we manually inspected the decompiled bytecode rather than relying on runtime behavior. Specifically, we obtained the bytecode for all sampled contracts from the Ethereum public dataset's `contracts' table. Subsequently, we employed the Panoramix\footnote{\url{https://github.com/palkeo/panoramix}} tool for bytecode decompilation. In identifying proxy contracts, we first focused on the presence of the \texttt{delegatecall} statement, a distinctive characteristic of proxies (Section \ref{sec:background}). Contracts lacking this statement were categorized as other, resulting in 290 such contracts. For the remaining 95 sample contracts, we analyzed the \texttt{delegatecall} statement to determine if the function being called via delegation exhibited a similar function selector to the function where the \texttt{delegatecall} statement was originally found in the sample contract. We accomplished this by examining the use of the first four bytes (i.e., the function selector) of the \texttt{calldata}. Contracts where this condition was met were labeled as proxy (90 instances) while those that did not meet this criterion were labeled as other (5 instances). Listing \ref{listing:eval-proxy-bytecode} shows an example of a proxy contract's bytecode\footnote{\url{https://etherscan.io/bytecode-decompiler?a=0x7f2722741F997c63133e656a70aE5Ae0614aD7f5}}. Line 5 shows the presence of the {delegatecall} statement, while line 6 shows that the call uses the first four bytes of the input \texttt{calldata}. Our ground truth contains 90 (23.4\%) proxy contracts and 295 (76.6\%) other contracts. Further details about our ground truth dataset can be found in our supplementary material package.

\noindent
\begin{figure}[t]
\begin{lstlisting}[frame=single, style=sidebyside, language=Solidity, caption={An example of a proxy contract's decompiled bytecode.},label={listing:eval-proxy-bytecode}]
def storage:
  stor0 is addr at storage 0

def _fallback() payable: # default function
  delegate stor0 with:  # 1) delegatecall
     funct call.data[0 len 4] # 2) similar interface 
     gas gas_remaining wei
     args call.data[4 len calldata.size - 4]
  if not delegate.return_code:
      revert with ext_call.return_data[0 len return_data.size]
  return ext_call.return_data[0 len return_data.size]
\end{lstlisting}
\end{figure}

We applied our proxy detection method to our ground truth dataset and evaluated its accuracy using precision, recall, and the F1-measure (Table \ref{tab:eval-01}). Our approach achieved perfect precision (100\%), but its recall was 68.1\%, indicating that we missed detecting 28 out of 90 proxies. A closer look at these cases revealed a common pattern: In 26 cases, the contract's source code indicated the presence of a proxy contract, but it had not been involved in any transactions. In other words, the proxy had never received any subsequent transactions since its creation. In the remaining two cases, although they received one more transaction each, these transactions did not trigger the proxy's functionality. Since our proxy detection method relies on proxies' runtime behavior, these cases went undetected. The F1-measure for proxy detection was 81.6\%, and for non-proxy cases, it was 95.5\%. It is worth noting that our approach demonstrated 100\% precision and recall in detecting active proxy contracts.

We conducted a comparative analysis of our detection approach with a related study by Salehi et al.~\citep{Salehi22}. \citeauthor{Salehi22} identified 1,427,215 proxies between Sep-05-2020 and Jul-20-2021. In contrast, our approach detected 1,723,309 proxies within the same time frame, marking a difference of 296,094 additional proxies detected by our method. Unfortunately, due to the unavailability of Salehi et al.'s proxy dataset, we couldn't perform a direct one-to-one comparison. Both approaches rely on behavioral patterns of contracts for proxy detection, meaning neither can identify inactive proxies. Nonetheless, our previous evaluation, based on a ground truth dataset of 385 contracts and the high precision achieved, instills confidence in the effectiveness of our approach.

\begin{table}[t]
\renewcommand{\arraystretch}{1.2}
\centering
\caption{The performance of our proxy detection approach.}
\begin{tabular}{llll}
\hline
               & \textbf{Precision} & \textbf{Recall} & \textbf{F1-measure} \\ \hline
\textbf{Proxy} & 100\%               & 68.9\%            & 81.6\%          \\
\textbf{Other} & 91.3\%                 & 100\%             & 95.5\%        \\ \hline
\end{tabular}
\label{tab:eval-01}
\end{table}

\begin{footnotesize}
    \begin{mybox}{Summary}
        \textbf{Proxy Detection Approach \& Evaluation}
        \tcblower We introduced an approach to identify proxy contracts based on their runtime behavior. Our evaluation, using a ground truth dataset of active proxy contracts, shows that our method achieves perfect precision with a 68.9\% recall rate. Our focus in this study was on active proxy contracts, those with at least one instance of proxy functionality usage, resulting in both perfect precision and recall for our proposed approach.
    \end{mybox}
\end{footnotesize}



\section{RQ1: How prevalent is the proxy mechanism in the Ethereum ecosystem?} \label{sec:rq1}

\noindent \textbf{Motivation.} 
 In our initial investigation (Section \ref{Sec:prem}), we observed a growing trend of transactions involving multiple contracts, with proxy contracts playing a central role. To gain deeper insights into the usage of the proxy pattern in the blockchain context, we now conduct a comprehensive study to track its prevalence since the inception of Ethereum.

\smallskip \noindent \textbf{Approach.} 
To detect proxies, we analyze the runtime behavior of over 50 million smart contracts collected in Section \ref{sec:data-collection}, using the method outlined in Section \ref{sec:proxy-detection}. After identifying the general prevalence of proxy contract instances, we conducted three separate analyses to examine their prevalence from different perspectives.

\begin{itemize}[label=\textbullet, itemsep = 3pt, topsep = 0pt]

    \item \textbf{Activity level viewpoint}. The goal of this analysis is to determine the activity level of proxy contracts. We also compared the activity levels of proxy contracts with non-proxy contracts (i.e., any contract that is not a proxy). We considered the number of inbound transactions to measure contract activity level. Then, we used the Complementary Cumulative Distribution Function (CCDF) to compare the number of inbound transactions per proxy contract and non-proxy contract as of Sep. 2022. In a CCDF graph, the y-axis represents the percentage of data points at or above a specific value, while the x-axis represents the variable of interest, in this case, the Number of Inbound Transactions. To further assess if the distribution of the number of inbound transactions for the proxy category is indeed greater than the latter non-proxy type, we employed the non-parametric one-tailed Mann–Whitney U test ($\alpha = 0.05$) and assessed the effect size using Cliff's Delta ($\delta$). The interpretation of $\delta$ is based on thresholds from \citep{Romano06}: \textit{negligible} if $|\delta| \le 0.147$, \textit{small} for $0.147 < |\delta| \leq 0.33$, \textit{medium} for $0.33 < |\delta| \leq 0.474$, and \textit{large} for greater values.

    \item \textbf{Usage context viewpoint}.The goal of this analysis is to determine the distinct usage contexts in which proxy contracts are utilized. More specifically, we analyzed the monthly count of such usage contexts. To identify distinct usage contexts, we performed the following seven steps: \textbf{i)} for each detected proxy contract, we extracted its deployer address, also known as the caller address of the proxy creation trace (e.g., \textit{C1} in Figure \ref{fig:trace-table}). This deployer address signifies the entity (e.g., either a developer or a smart contract on behalf of the developer) responsible for deploying the proxy contract. \textbf{ii)} We collected the proxy's associated logic contracts. This inclusion is necessary because developers, even within the same organization, may clone proxy contracts for different purposes. To account for these variations in purpose, we incorporated the addresses of the logic contracts to which proxy contracts have ever delegated calls since the logic contract defines the underlying logic behind the proxy. \textbf{iii)} using the proxy address, proxy bytecode, logic contracts, and deployer address, we clustered proxy contracts based on their bytecode and deployer address. This allowed us to identify similar proxies created by the same entity. \textbf{iv)} within each cluster, we constructed a graph where the nodes represent proxy contract addresses (proxy nodes) and their associated logic contract addresses (logic nodes). An edge exists between a proxy node and a logic node if the proxy delegates calls to the logic contract. \textbf{v)} we identified connected components within the graph. Each component represents a unique usage context, representing a set of similar proxy contracts deployed by a single entity for a similar purpose. \textbf{vi)} within each context, we retained the oldest proxy contract, recorded its creation timestamp as the starting time of the context, and excluded any other duplicates. \textbf{vii)} finally, we computed the monthly count of contexts.

    \item \textbf{Stakeholder adoption viewpoint}. The goal of this analysis is to determine the engagement of EOAs (a.k.a., stakeholders) in creating proxy contracts. We analyzed the count and the ratio of EOAs who initiated a proxy creation transaction as of each month. An EOA can either directly deploy a proxy or a proxy can be deployed upon the EOA's request to a third-party contract (e.g., a proxy factory contract). In either case, the EOA initiates a creation transaction through which we can identify its unique address. More specifically, for a given $i^{th}$ month, we counted the total number of EOAs who initiated a proxy contract creation transaction between Aug-07-2015 to the end of $i^{th}$ month. Additionally, we collected the general number of EOAs who initiated any contract creation transaction as of the $i^{th}$ month. Then, we computed the ratio of EOAs who initiated a proxy contract as of each month by dividing the former figure by the latter one. 
    
    
    \item \textbf{Proxy utilization viewpoint}. The goal of this analysis is to shed light on the usage of proxy contracts through analyzing smart contract transactions. While we showed earlier that 98.4\% of a sample of multi-call transactions with a \texttt{delegatecall} involves a proxy contract (Section\ref{Sec:prem}), we did not investigate how proxy usage evolved over time due to limitations in applying Etherscan's Proxy Verification at scale. Our proposed proxy detection approach overcomes this issue. Hence, we set out to conduct this evolutionary analysis in this RQ. We first studied the monthly ratio of transactions in which proxies are used. Next, focusing on the monthly ratio of multi-call transactions (Section\ref{Sec:prem}), we calculated a new monthly ratio of transactions in which at least one proxy participated. Since transactions represent how contracts interact, this metric highlights the extent to which proxies are being used in the design of contracts.
\end{itemize}

\begin{figure}[t]
\begin{minipage}[t]{0.49\linewidth} 
 \includegraphics[width=\linewidth]{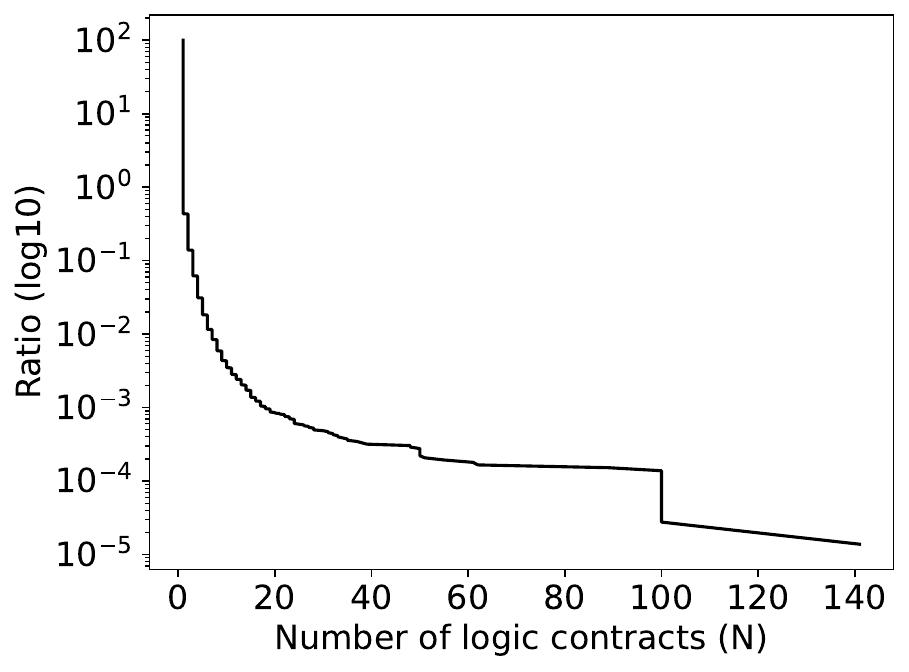}
 \caption{The CCDFs of the number of inbound transactions per proxy and non-proxy contracts.}
 \label{fig:rq1-ccdf-number-of-logics-per-proxy}
\end{minipage}\hfill 
\begin{minipage}[t]{0.49\linewidth}
 \includegraphics[width=\linewidth]{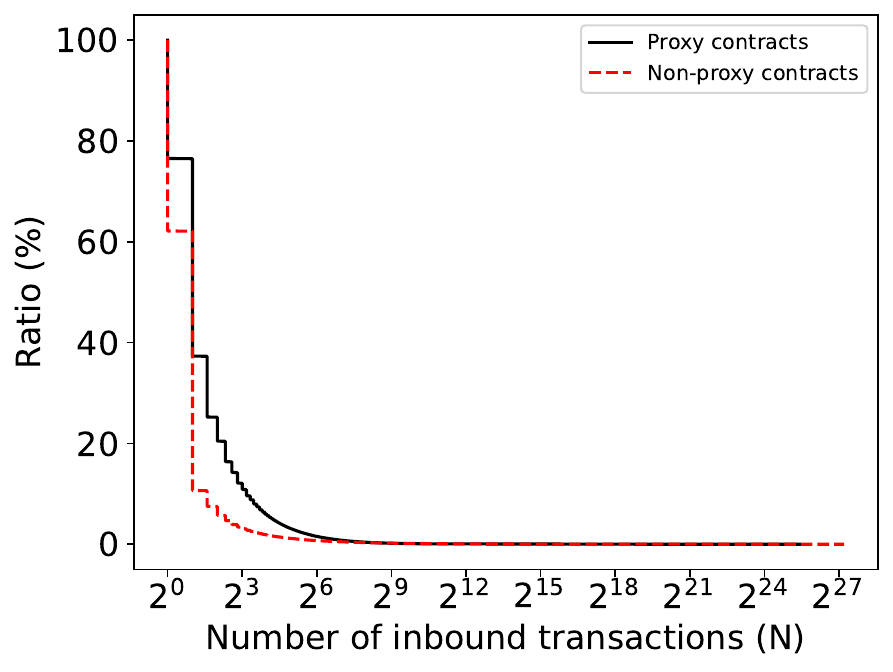}
 \caption{The monthly number of distinct usage contexts.}
 \label{fig:rq1-activity-level}
\end{minipage}
\end{figure}

\smallskip \noindent \textbf{Findings.}\observation{Over 14\% of all instantiated smart contracts are proxies.} \label{obs:rq1-01} 50,845,833 contracts were deployed to Ethereum as of Sep. 2022. Out of these, 7,241,339 (14.2\%) are active proxy contracts. Each such proxy is paired with at least one logic contract. A proxy can be associated with multiple logic contracts throughout its life, adjusting for contract upgrades and evolving needs. Some proxies also function as routers, directing requests to the appropriate logic contract, as seen in setups like the Diamond Proxy Pattern \citep{EIP-2535} where they interact with several facet contracts. We found 7,296,032 pairs of proxy and logic contracts. Figure \ref{fig:rq1-ccdf-number-of-logics-per-proxy} depicts the CCDF of the number of logic contracts that a proxy has ever interfaced with per proxy contract. Analysis of the graph reveals that the vast majority of proxy contracts (99.56\%) are linked to a single logic contract. Furthermore, as we progress towards the tail end of the distribution, the probability of proxy contracts having multiple logic contracts experiences a sharp decline, reaching below $10^{-4}$\% at N = 100, where it undergoes a sudden dip that continues until the end of the distribution. In the extreme case, the proxy contract with 141 logic contracts is BZxProxy\footnote{\url{https://etherscan.io/address/0x1cf226e9413addaf22412a2e182f9c0de44af002}} owned by the bZx Protocol\footnote{\url{https://b0x.network/}}. This contract holds a mapping of services to target logic contracts where each service is implemented. Then, depending on the requested service it delegates the call to a proper logic contract. Therefore, it acts as a router.

\smallskip \observation{[ACTIVITY LEVEL] The distribution of the number of inbound transactions per proxy is statistically greater than the distribution underlying non-proxy contracts.}\label{obs:rq1-activity-level} Figure \ref{fig:rq1-activity-level} illustrates the CCDF of the number of inbound transactions per proxy contract, represented by the black line. The graph shows a long tail, indicating that a small proportion of proxy contracts receive many transactions, while most have very few transactions. When comparing the CCDFs of the proxy and non-proxy contracts, it is evident that proxy contracts are more likely to have a higher number of inbound transactions (N) compared to non-proxy contracts for 1 <= N <= 256 (initial range). However, for larger values of transaction counts (N > 256), non-proxy contracts start to surpass proxy contracts in terms of likelihood, although it is important to note that the proportion of contracts with a large number of inbound transactions in this range remains relatively small for both proxy and non-proxy contracts. In the extreme tail of the distribution (i.e., outliers), it becomes evident that the highest number of inbound transactions is associated with non-proxy contracts. A one-tailed Mann-Whitney test confirms this distributional difference as statistically significant (p-value < 0.05). The Cliff's Delta calculation further reveals a medium effect size ($\delta$ = 0.34). Thus, proxy contracts are utilized more than non-proxy contracts. This underscores the paramount importance of proxy contracts.

\begin{figure}[!t]
\begin{minipage}[t]{0.49\columnwidth}
 \includegraphics[width=\linewidth]{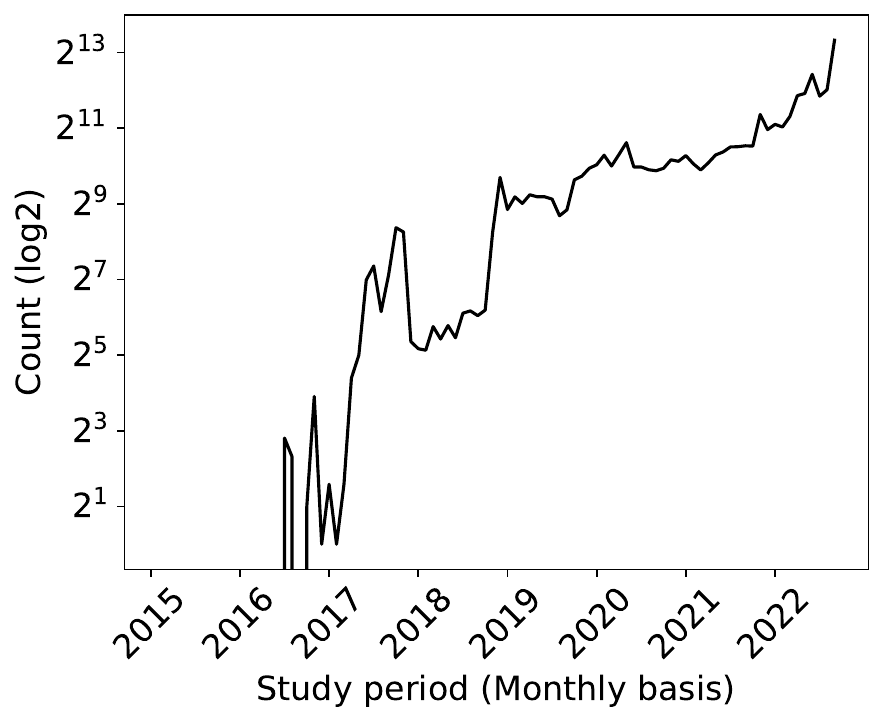}
 \caption{The monthly number of distinct usage contexts.}
  \label{fig:rq1-usage-context}
\end{minipage}\hfill 
\begin{minipage}[t]{0.49\columnwidth}
 \includegraphics[width=\linewidth]{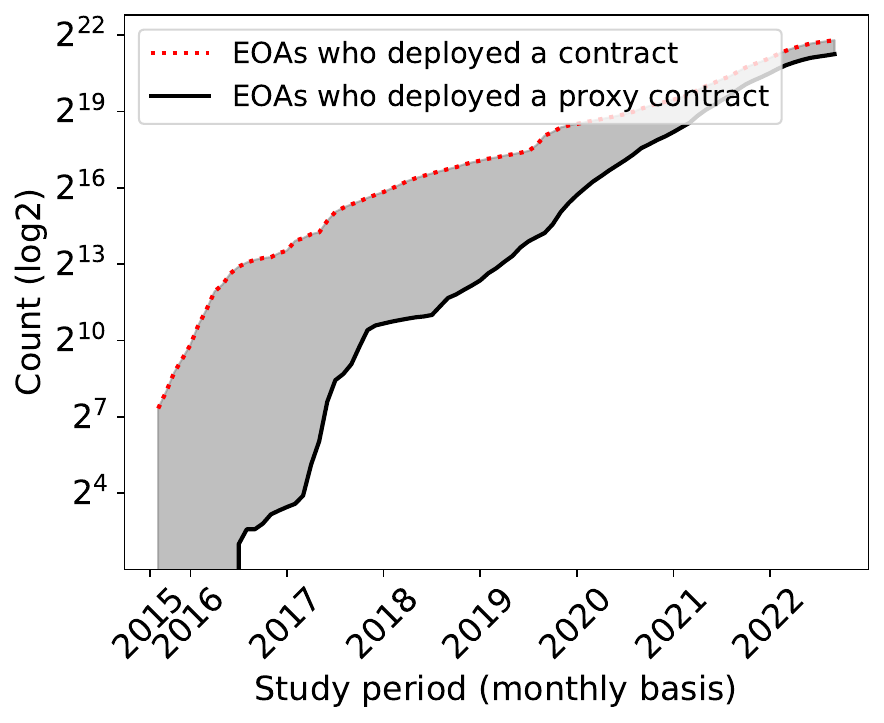}
 \caption{The number of EOAs instantiated a (proxy) contract as of each month.}
  \label{fig:rq1-number-of-eoas-who-initiated-a-proxy}
\end{minipage}
\end{figure}

\smallskip \observation{[USAGE CONTEXT] There is a clear increasing trend in the creation of distinct usage contexts in practice, reaching just below 80K distinct contexts by Sep. 2022.}\label{obs:rq1-02} Figure~\ref{fig:rq1-usage-context} illustrates a noticeable trend of increasing usage contexts over the years, with a significant surge in activity observed from 2017 onwards. While the initial years (2015-2016) show minimal activity, subsequent years demonstrate consistent growth, reaching its peak in Aug. 2022 with over $2^{13}$ usage contexts. Notable fluctuations occur on a monthly basis, with certain months experiencing exceptionally high levels of usage. Indeed, since each distinct usage context is a different use case, we can claim that proxies are becoming increasingly popular in different contexts. In addition, the graph reveals that the idea of the proxy pattern has been around since the early days of Ethereum. The first usage context for proxy contracts emerged in mid-2016, just 10 months after Ethereum went live. Finally, we identified 79,171 distinct usage contexts over the studied period.

\begin{figure}[!t]
\begin{minipage}[t]{0.49\columnwidth}
 \includegraphics[width=\linewidth]{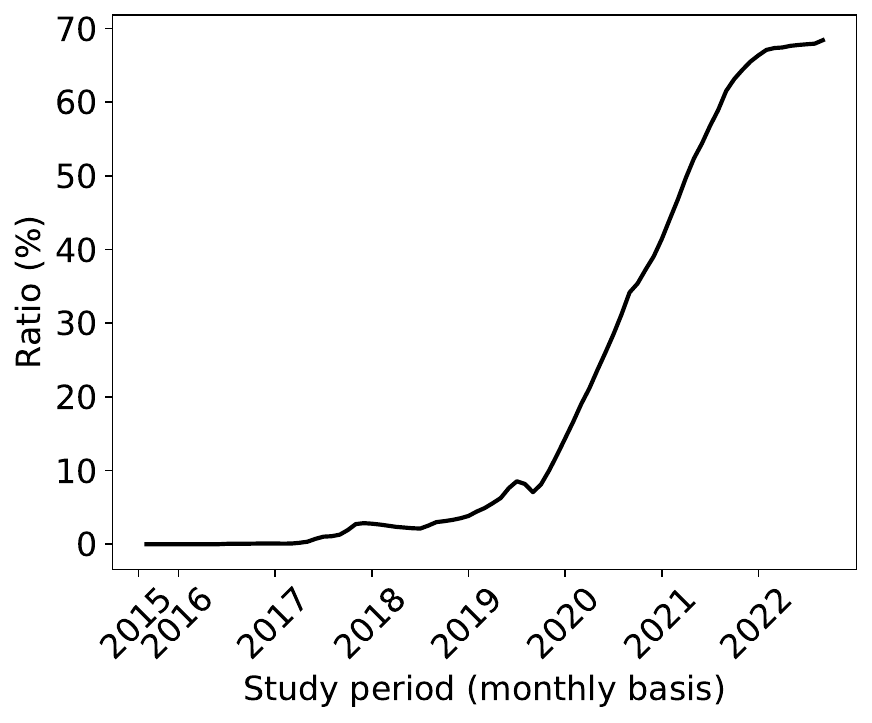}
 \caption{The ratio of EOAs instantiated a proxy contract as of each month.}
\label{fig:rq1-ratio-of-eoas-who-initiated-a-proxy}

\end{minipage}\hfill 
\begin{minipage}[t]{0.49\columnwidth}
 \includegraphics[width=\linewidth]{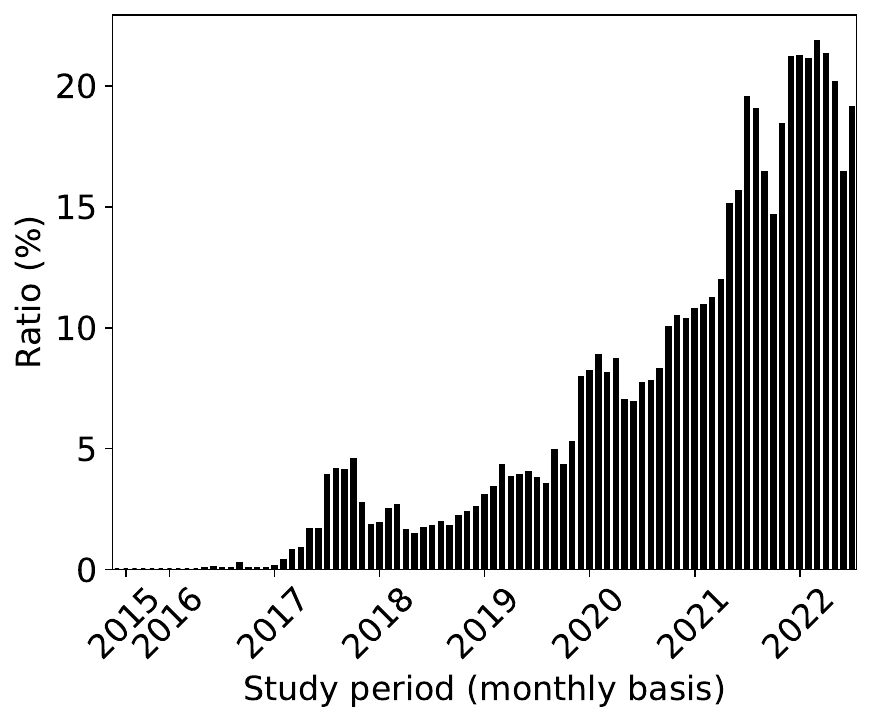}
 \caption{The monthly ratio of transactions that involve at least one proxy contract.}
  \label{fig:monthly-ratio-of-transactions-with-proxy}
\end{minipage}
\end{figure}

\smallskip \observation{[STAKEHOLDER ADOPTION] There is an upward trend in the number of EOAs who initiated a proxy contract transaction, surpassing two-thirds of all EOAs who instantiated any contract in 2022.}\label{obs:rq1-stakeholder-adoption} As evidenced by Figure \ref{fig:rq1-number-of-eoas-who-initiated-a-proxy}, the number of EOAs that instantiated at least one contract (the red line) has gradually increased each month, peaking at 3,636,192 at the end of the study period. In addition, the figure implies that EOAs started employing proxy contracts in mid-2016 (black line), when the first proxy contract was instantiated (Observation \ref{obs:rq1-02}). Since then, this figure has witnessed rapid growth, peaking at 2,487,994 as of Sep. 2022. The gray area between the red and black lines shows that, while the gap between the number of EOAs who instantiated a smart contract and those who instantiated an instance of a proxy contract was large at the beginning, the area significantly narrowed down towards the end of the study period, indicating a higher tendency for instantiating proxy contracts among the EOAs. Additionally, Figure \ref{fig:rq1-ratio-of-eoas-who-initiated-a-proxy} shows that the ratio of EOAs initiated a proxy contracts transaction has been more or less stable with marginal growth until mid-2019. After this, the figure experienced notable linear growth up to 68.4\%, where it plateaued at the end of the study period. This implies that as of September 2022, at least one proxy contract was instantiated by 68.4\% of EOAs that instantiated smart contracts.

\begin{figure}[!t]
\centering
\begin{minipage}[t]{0.49\columnwidth}
 \includegraphics[width=\linewidth]{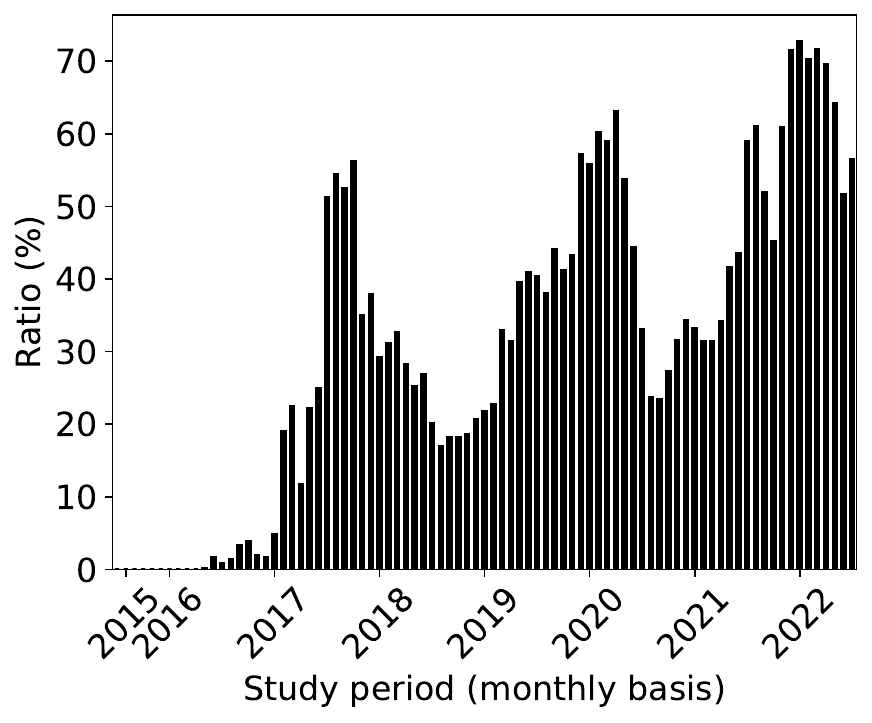}
 \caption{The monthly ratio of multi-contract transactions that involve at least one proxy contract.}
\label{fig:monthly-ratio-of-multi-contract-transactions-with-proxy}
\end{minipage}\hfill
\end{figure}

\smallskip \observation{[PROXY UTILIZATION] Over 20\% of transactions in 2022 involve a proxy contract. }\label{obs:rq1-04} Figure \ref{fig:monthly-ratio-of-transactions-with-proxy} shows the monthly ratio of transactions in which a proxy played a role. This indicates that, from a behavioral perspective, the role of proxies in carrying out smart contracts' functionalities has become more important over time. For instance, proxies were involved in over 20\% of all transactions that occurred after 2022. The role of proxies becomes even bolder when applications' designs become more fine-grained and modularized. Figure \ref{fig:monthly-ratio-of-multi-contract-transactions-with-proxy} indicates the monthly ratio of multi-call transactions in which a proxy contract participated. The graph shows that despite fluctuation, proxy contracts are becoming remarkably critical in the modular era of software with almost 70\% of multi-call transactions using at least one proxy contract in 2022.

\begin{footnotesize}
    \begin{mybox}{Summary}
        \textbf{RQ1: \RQOne}
        \tcblower
        Our assessment from the four viewpoints of activity level, usage context, stakeholder adoption, and proxy utilization shows that the tendency to use proxy contracts is growing. We also confirm the substantial relevance of proxies in modular application designs.
    \end{mybox}
\end{footnotesize}

\section{RQ2: \RQTwo}\label{sec:rq2}


\smallskip \textbf{Motivation.} In Observation \ref{obs:rq1-01}, we found that over 7.2M (14\%) of all deployed contracts are proxies. Given this high ratio, we wish to understand how proxies are deployed into DApps in practice. Such an investigation should provide insights into the design of modern modular DApps.

\smallskip \noindent \textbf{Approach.} To identify various patterns of deploying proxies, we conducted a mining process to trace the deployment infrastructure involved. 
In particular, for each proxy, we accessed its creation transaction and examined its trace table (Section \ref{sec:background}) to pinpoint the trace relevant to the proxy's creation. This specific trace had to meet two criteria: its type had to be ``create'', and the callee address had to match the proxy contract's address (step 1). We then retrieved the trace's caller address, which we referred to as the ``deployer address'' (step 2). Subsequently, we determined whether the deployer address belonged to an EOA or a CA (step 3). We accomplished this by cross-referencing the deployer address with the list of all CAs, obtained by querying the ``contracts'' table from the Ethereum public dataset. If the deployer address belonged to a CA, we recorded it and iteratively applied the same process to the CA until we identified the root EOA who is in charge of deploying the first contract of the proxy deployment infrastructure (step 4). Alternatively, if the deployer address belonged to an EOA, we recorded it as the entity that directly deployed the proxy contract (step 5). Once the root EOA was identified, we established a sequence of contract creations, starting with a root EOA address and concluding with the proxy contract address. 

\begin{figure}
\centering

    \subfigure[The sequence of contract creations.]{\includegraphics[width=0.475\linewidth]{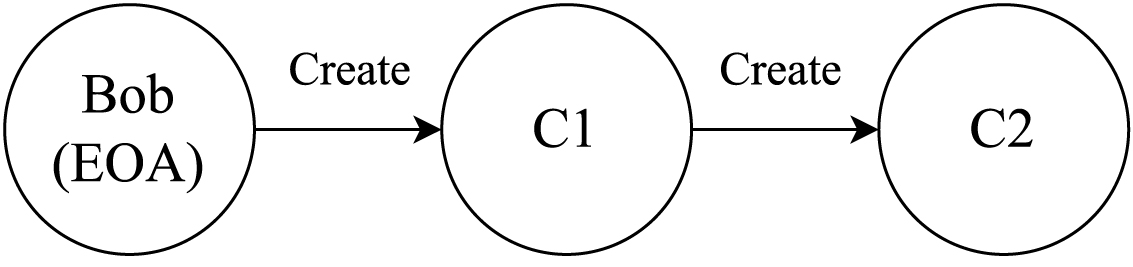}\label{fig:rq2-seq-of-contract-creation}}
    \hfill
    \subfigure[The labeled sequence of contract creations.]{\includegraphics[width=0.475\linewidth]{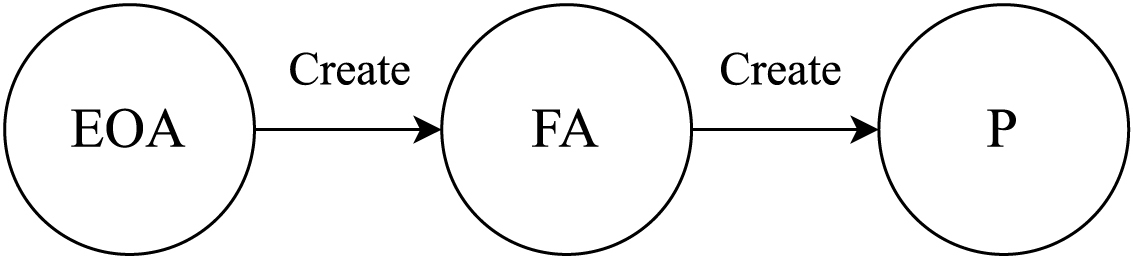}\label{fig:rq2-proxy-creational-pattern}}
    
\caption{A sequence of contract creations and its labeled version.}

\end{figure}

Finally, to extract patterns, we assigned labels to all nodes within a given sequence (step 6). Specifically, the initial node of every sequence, being a root EOA, was designated as such. An intermediary node was labeled as a factory (FA), indicating its role in spawning another contract during runtime. If the intermediary factory node also represented a proxy contract, it was labeled as a proxy factory (PF). Proxy factories typically wrap a factory contract and are used to either efficiently clone the factory contract's functionality or enable upgrades to the factory. The concluding node of the sequence was consistently labeled as a proxy (P). This sequence of labeled nodes illustrated how a proxy was deployed, referred to as a ``proxy creational pattern''. For example, consider the scenario where Bob (the root EOA) created a factory contract C1, and C1, in turn, created a proxy contract C2. Figure \ref{fig:rq2-seq-of-contract-creation} and \ref{fig:rq2-proxy-creational-pattern} depict the sequence of contract creations and its labeled version for this example.

After obtaining the creational patterns, we performed a series of analyses. Initially, we examined the distinct proxy creational patterns and developed a metamodel using a UML class diagram. This metamodel succinctly communicates the various methods employed for proxy deployment. Subsequently, we analyzed the extent to which each creational pattern was used in different contexts. We used our method for detecting usage contexts (Section \ref{sec:rq1}) and applied it to identify different usage contexts under each creational pattern. 


Furthermore, we conducted a comparison between off-chain and on-chain deployment styles regarding the number of deployed proxy contracts per usage context. For each of the 79,171 identified usage contexts (Section \ref{sec:rq1}), we categorized them as either off-chain or on-chain based on the deployment style used for their proxy contracts and computed their size in terms of the number of proxy contracts. Subsequently, we performed a comparison of the CCDFs of the number of proxy contracts per usage context between these two styles. Additionally, a one-tailed Mann–Whitney U test ($\alpha = 0.05$) was employed to evaluate whether the distribution of the number of proxy contracts per usage context is greater for the on-chain style compared to the off-chain style. Furthermore, we extended this analysis to the creational pattern level, focusing on the top-5 on-chain patterns (in terms of proxy contract count) and the sole off-chain pattern. We applied the same aforementioned methodology to compare on-chain and off-chain patterns. Finally, we categorized usage contexts under each style based on their size into either singleton (N=1) or non-singleton (N>1). Then, we performed a chi-square test to assess if there is any relationship between the size of usage contexts and the deployment style. We measured the effect size magnitude using the standard Phi ($\phi$) measure and interpreted it as follows: \textit{small} if $0.0 < |\phi| \leq 0.3$, \textit{medium} for $0.3 < |\phi| \leq 0.5$, and \textit{large} for greater values.

\begin{figure}[!t]
\centering
 \includegraphics[width=0.50\columnwidth,]{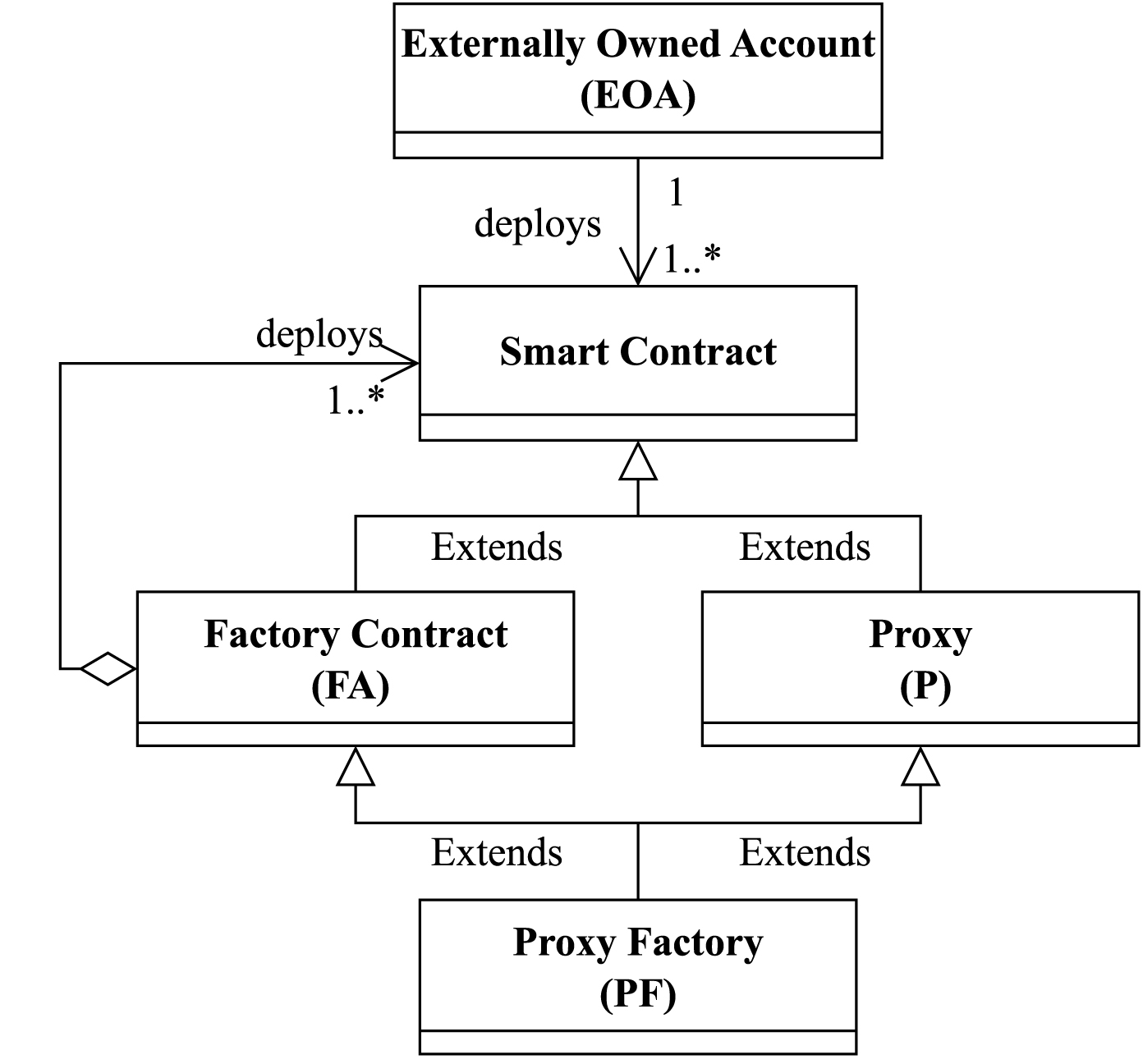}
  \caption{A metamodel that summarizes our proxy creational patterns.}
  \label{fig:rq2-01}
\end{figure}

\smallskip \noindent \textbf{Findings.} \observation{We found 12 different creational patterns for deploying proxy contracts.}\label{obs:rq2-introduce-12-patterns} The 7.2M proxies are deployed through 12 creational patterns. The creational patterns in Table \ref{tab:rq2-01} read from left to right. The ``>'' operator shows a deployment relationship where the left and right operands represent the deployer and deployee type in a pattern, respectively. Figure \ref{fig:rq2-01} shows a metamodel that summarizes the 12 detected creational patterns using a UML class diagram. All patterns are initiated by an EOA, since Ethereum transactions can only be initiated by an EOA (Section~\ref{sec:background}). More specifically, an EOA can either directly deploy instances of \textit{proxy contracts} (P) or deploy a \textit{factory contract} (FA). A factory contract is able to deploy (clones of) a certain template contract at runtime. Inspired by the composite design pattern, the metamodel shows that a factory (i.e., the composite) can deploy other factory contracts or proxy contracts (i.e., the leaf). Furthermore, there is another class of factory contract, namely \textit{proxy factory} (PF), where the factory itself is a proxy too. This type of factory can enable developers to create upgradeable factories, allowing them to alter factories' behavior and configuration (e.g., changing the hard-coded template contract) if deemed necessary.

\begin{table}[t]
\centering
\caption{Overview of the 12 detected proxy contract creational patterns sorted based on the category and the proxy instance count.}
\label{tab:rq2-01}
\resizebox{\textwidth}{!}{
\begin{tabular}{@{}cllrr@{}} 
\toprule
\textbf{Id} & \textbf{Creational Pattern} & \textbf{Category} & \multicolumn{1}{c}{\textbf{Usage Context Count}} & \multicolumn{1}{c}{\textbf{Proxy Instance Count}} \\ 
\midrule
1  & EOA > P                   & Off-chain & 30,933 (39.07\%)      & 50,174 (0.69\%)   \\
2  & EOA > FA > P              & On-chain  & 31,520 (39.81\%)      & 6,618,012 (91.39\%) \\
3  & EOA > PF > P              & On-chain  & 565 (0.71\%)          & 379,293 (5.24\%)  \\
4  & EOA > FA > FA > P         & On-chain  & 182 (0.23\%)          & 123,349 (1.70\%)  \\
5  & EOA > FA > FA > FA > FA > P & On-chain  & 7 (0\%)              & 35,764 (0.49\%)   \\
6  & EOA > FA > PF > P         & On-chain  & 15,627 (19.74\%)      & 31,586 (0.44\%)   \\
7  & EOA > PF > PF > P         & On-chain  & 138 (0.17\%)          & 1,968 (0.03\%)    \\
8  & EOA > FA > PF > PF > P    & On-chain  & 183 (0.23\%)          & 589 (0.01\%)      \\
9  & EOA > FA > FA > FA > P    & On-chain  & 11 (0.01\%)           & 599 (0.01\%)      \\
10 & EOA > FA > FA > PF > P    & On-chain  & 3 (0.01\%)            & 3 (0.00\%)        \\
11 & EOA > FA > FA > PF > PF > P & On-chain  & 1 (0\%)              & 1 (0.00\%)        \\
12 & EOA > FA > FA > PF > PF > PF > P & On-chain  & 1 (0\%)          & 1 (0.00\%)        \\
\bottomrule
\end{tabular}}
\end{table}

\smallskip \observation{The 12 proxy creational patterns can be classified into two major styles: i) on-chain and ii) off-chain deployment.}\label{obs:rq2-introduce-metamodel} On-chain deployment is a style through which proxy instances are deployed by smart contracts at runtime, covering 11 of the patterns. On the other hand, off-chain deployment is a style in which proxy instances are deployed by an EOA through deployment scripts that are maintained \textit{outside} the blockchain. \textit{While proxy instances of both styles are eventually deployed to the Ethereum blockchain, the main difference between these two styles is concerned with where their deployment infrastructure is operating}. In the off-chain deployment, scripts for deploying proxies operate outside the blockchain in a centralized fashion. On the other hand, in the on-chain deployment style, proxy instance creation is performed by smart contracts that operate on the blockchain in a decentralized fashion. The latter ensures that every instantiation of a proxy (and the way in which this happens) will always be traceable and recorded in the blockchain. Therefore, despite the increased complexity of on-chain deployment, such a style provides better traceability. See Sections~\ref{subsec:benefits-and-drawbacks-of-on-chain}, \ref{subsec:benefits-and-drawbacks-of-off-chain} and Appendix \ref{subsec:gas-cost-complexity-of-dep-style} for a detailed discussion of these two deployment styles.


\smallskip \observation{There is a significantly higher diversity of deployment scenarios, as evidenced by the wider range of usage contexts for the on-chain style (60.9\%) than the off-chain style (39.1\%).}\label{obs:rq2-dep-style-popularity} When assessing deployment scenarios and their diversity, an essential metric to consider is the ratio of usage contexts (Section \ref{sec:rq1}). Our analysis reveals a noteworthy dichotomy between on-chain and off-chain categories. Specifically, a majority, encompassing 60.9\% of the examined usage contexts, aligns with the on-chain paradigm. In contrast, 30.1\% are affiliated with off-chain categories. This distribution emphasizes the prominence of on-chain deployments within the blockchain ecosystem. The two most prevalent patterns in terms of usage contexts are \textit{EOA > FA > P}, accounting for 39.81\% of the observed contexts, closely followed by \textit{EOA > P}, which comprises 39.07\%.



\smallskip \observation{The majority of proxy instances (99.3\%) are systematically deployed via the on-chain style.}\label{obs:rq2-dep-style-proxy-count} 7,191,165 (99.3\%) proxies are deployed using the 11 on-chain creational patterns, whereas the 50,174 (0.69\%) proxies that are deployed using the off-chain style. Out of 11 the on-chain creational patterns, the second (i.e., \textit{EOA > FA > P}) and third (\textit{EOA > PF > P}) patterns are the top-2 most used ones in terms of the number of deployed proxy contracts. The remaining 9 on-chain patterns are a specialization of the third and second patterns, albeit they are longer. While longer patterns advocate higher reusability and interoperability among contracts, we observe that simpler patterns are typically more common in terms of both usage context count metric and proxy instance count. As the top-5 on-chain patterns comprising 99.26\% of deployed proxy contracts, we systematically identified and analyzed a large DApp for each of them (see Appendix \ref{apx:discussion-of-largest-dapps-for-the-top-5-patterns} for examples of the usage of longer on-chain patterns in DApps).

\begin{figure}[t]
\begin{minipage}[t]{0.49\columnwidth}
 \includegraphics[width=\linewidth]{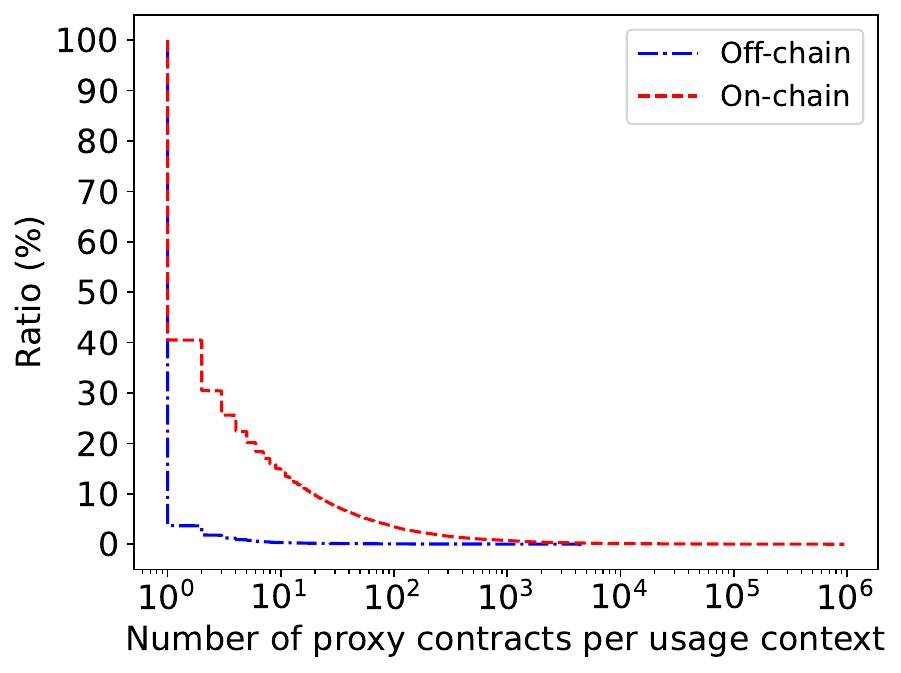}
 \caption{CCDFs of the number of proxy contracts per usage context of the on-chain and off-chain deployment style.}
\label{fig:ccdf-context-size-per-style}
\end{minipage}\hfill 
\begin{minipage}[t]{0.49\columnwidth}
 \includegraphics[width=\linewidth]{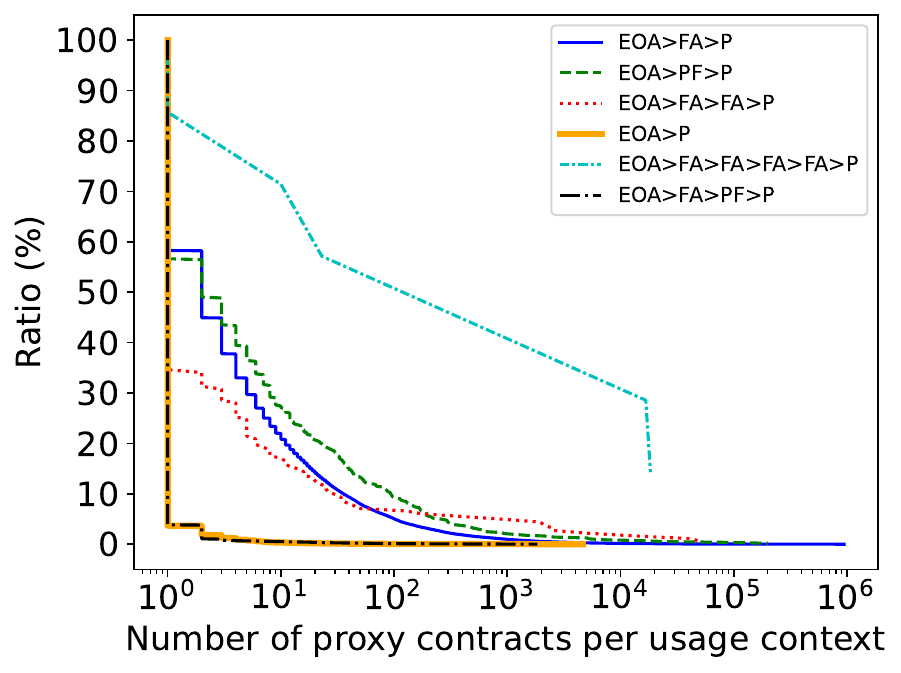}
 \caption{CCDFs of the number of proxy contracts per usage context of top-5 on-chain and off-chain patterns.}
\label{fig:ccdf-context-size-per-top-6-patterns}
\end{minipage}
\end{figure}

\smallskip \observation{On-chain deployment styles have larger usage contexts compared to the off-chain style.}\label{obs:rq2-dep-style-and-contet-size} Figure \ref{fig:ccdf-context-size-per-style} shows our comparison of deployment styles. It is evident that the CCDF of on-chain styles constantly stays above that of the off-chain style, indicating that the likelihood of observing larger usage contexts is consistently higher for the on-chain style. The Mann–Whitney U test shows that the distribution of the size of usage contexts for the on-chain style is greater than that of the off-chain style (p-value < 0.05) with Cliff's Delta showing a medium effect size ($\delta = 0.37$). Finally, Figure \ref{fig:ccdf-context-size-per-top-6-patterns} compares the CCDF for top-5 on-chain patterns and the off-chain ones. It is evident that on-chain patterns typically exhibit a higher probability when compared to the off-chain pattern (i.e., red line). Furthermore, the CCDF curve of on-chain patterns has a slower decrease rate. This indicates that larger usage contexts (i.e., those with more proxy contracts) in on-chain patterns are more frequently encountered. In contrast, the only off-chain pattern (i.e., EOA > P) displays a CCDF curve that drops more rapidly, indicating a lower probability of encountering higher values within its distribution. Finally, Table \ref{tab:number-of-usage-contexts-per-deployment-style} compares the number of singleton usage contexts and non-singleton per deployment style. In particular, we found that the likelihood that usage contexts of the on-chain style will have more than one proxy contract is approximately ten times higher than that of the off-chain style (40.4\% versus 3.7\%, respectively. The chi-square test revealed that there is a relationship between larger usage contexts and the choice of deployment style with a medium effect size (0.4).

\begin{footnotesize}
    \begin{mybox}{Summary}
        \textbf{RQ2: \RQTwo}
        \tcblower
        We found 12 creational patterns with a clear distinction between on-chain and off-chain deployment styles. The on-chain style, encompassing 11 of these patterns and accounting for 99.3\% of the proxy instances, is marked by its complexity and enhanced traceability. The on-chain \textit{EOA > FA > P} and the off-chain \textit{EOA > P} emerge as the most prominent, dominating both in terms of usage contexts. Notably, on-chain deployments are characterized by greater diversity and a higher probability of larger usage contexts, whereas the off-chain pattern is more likely to be used in smaller contexts.
    \end{mybox}
\end{footnotesize}

\begin{table}[t]
\centering
\caption{The number of usage contexts per deployment style and the usage context size.}
\label{tab:number-of-usage-contexts-per-deployment-style}
\begin{tabular}{@{}lrr@{}}
\toprule
\multicolumn{1}{c}{\textbf{Number of proxy contracts (N)}} & \multicolumn{2}{c}{\textbf{Deployment Style}} \\ 
\cmidrule(r){2-3}
 & \multicolumn{1}{c}{\textit{On-chain}} & \multicolumn{1}{c}{\textit{Off-chain}} \\ 
\midrule
N > 1 & 19,547 (40.5\%) & 1,140 (3.7\%) \\
N = 1 (Singleton) & 28,691 (59.5\%) & 29,793 (96.3\%) \\
\midrule
\textbf{Total} & 48,238 & 30,933 \\
\bottomrule
\end{tabular}
\end{table}



\section{RQ3: \RQThree}
\label{sec:rq3}

\noindent \textbf{Motivation.} The growing use of proxy contracts in the Ethereum network has sparked interest in understanding the different types of proxies and their implementation practices \citep{Salehi22, evm_proxy_detection}. 

Proxies can be classified into \textit{forwarders} and \textit{upgradeability} based on their purpose~\citep{eip_897, Salehi22}. 
Understanding the prevalence of these two proxy contract types can provide insights into the design choices made by DApp developers. Despite initial research being conducted in 2020-2021 on the classification of proxies as upgradeability and forwarders\citep{Salehi22}, there is a lack of empirical evidence regarding their current prevalence on Ethereum. This question seeks to fill this gap by examining the prevalence of proxy types and how they are created, ultimately shedding light on how different proxy types are typically deployed.

Furthermore, there exist several reference implementations for proxy contracts on Ethereum, such as the EIP-897 Delegate Proxy Pattern~\citep{eip_897}, the EIP-1967 Transparent Proxy Pattern~\citep{eip1967}, and the OpenZeppelin Proxy pattern~\citep{openzeppelin_2020_proxy_pattern} (Section~\ref{sec:background}). While these standards offer different features and trade-offs in terms of security, gas efficiency, flexibility, and functionality, they remain similar in terms of the common behavior of a proxy. Thus, our approach is able to detect them regardless of the standard. In this RQ, we study the prevalence of seven reference proxy implementations, such that we can determine their popularity and understand whether developers adhere to those implementations at all.

\smallskip \noindent \textbf{Approach.} In the following, we describe the three analyses that we perform:

\begin{itemize}[label=\textbullet, itemsep = 3pt, topsep = 0pt]
    \item \textbf{On the prevalence of forwarder and upgradeability proxies.} Literature has simply classified proxies into two major types based on their purpose~\citep{eip_897, Salehi22}: forwarders and upgradeability proxies. \citet{Salehi22} proposed an approach for automatically detecting upgradeability and forwarder proxies. Given that a replication package is not provided, we manually implemented their approach and applied it to a statistically representative sample of proxies. Given 79,171 detected usage contexts (Section \ref{obs:rq1-02}), we pick one representative proxy contract per usage context. Since all proxy contracts within a usage context are similar, the representative proxy contract is randomly picked. Subsequently, we used the set of representative proxy contracts while drawing samples. 
    In the first analysis, we studied the ratio of usage contexts that employed either upgradeability or forwarder proxy contracts. We drew a statistically representative sample of representative proxy contracts, accounting for 385 instances. The sample size was estimated based on a 95\% confidence level and confidence interval of 5. We then classify the samples into either upgradeability or forwarder categories. To ensure a consistent approach while manually classifying proxies into forwarder and upgradeability types, we established a protocol and precisely followed it. Initially, the first author of this paper extracted the classification process from the study of \citet{Salehi22} and summarized the steps. We explain this process in Appendix \ref{apx:proxy-classification-process}. Afterward, the first and second authors together applied the extracted approach to every proxy instance in the sample to examine its type. In case of disagreement, the third author mediated a discussion to resolve it (three cases). Also, if we could not determine the type of a proxy, we marked it as \textit{unknown} (zero cases). The manual classification took 17 man-hours. Our replication package includes the sample dataset for reference.

    \item \textbf{On the relationship between the proxy contract purpose and the type of creational pattern.} We aimed to analyze if there is any relationship between the purpose of proxy contracts and top-5 (in terms of the number of proxy contracts) creational patterns. We first placed every representative proxy contract into a corresponding creational pattern. Then, we picked 385 random proxy contracts across the top-5 patterns (i.e., ideally 77 instances per pattern). However, since the fifth pattern only has seven usage contexts and so does seven representative contracts, we distribute the remaining 70 instances of this pattern across the top-4 patterns. Afterwards, we manually categorized each sample proxy contract into either upgradeability vs forwarder types using a similar protocol as explained in the previous analysis. In two cases, the third author mediates to resolve the conflicts. The manual classification took 16 man-hours. Subsequently, we analyzed the trend of proxy types across the top-5 patterns. 

    \item \textbf{On the adherence of proxy contracts to known reference standards.} The goal of this analysis is to investigate the extent to which proxy contracts adhere to well-known proxy reference implementations. We used a tool called \textit{evm-proxy-identification}~\citep{evm_proxy_detection} that identifies seven known implementations of the proxy pattern including four ERC-897 \cite{eip_897}, ERC-1167~\citep{eip_1167}, ERC-1822~\citep{eip_1822}, ERC-1967~\citep{eip1967}, ERC-1967 Beacon~\citep{eip1967} proposals, OpenZeppelin Proxy Pattern~\citep{openzeppelin_2020_proxy_pattern}, and GnosisSafeProxy\footnote{\url{https://github.com/safe-global/safe-contracts}}. Each of these implementations, except ERC-1167, standardizes a unique known storage slot for storing the address of the logic contract. The \textit{evm-proxy-identification} tool examines contracts' storage slots to detect their types. For ERC-1167, the tool detects them based on their bytecode. Due to a limit on the number of API requests (quota), we applied the tool to a sample of representative proxies. However, this time we opted for a larger sample. More specifically, we randomly picked 13,776 proxies based on a sample size estimation with a 99\% confidence level and a confidence interval of 1. Subsequently, we applied the tool to categorize each proxy from the sample into one of our seven categories. We also added another category namely ``Customized'' for the cases where a proxy contract does not match any of the reference implementations. Our replication package includes the sample dataset for reference.

\end{itemize}

\smallskip \noindent \textbf{Findings.} \observation{Just above two-thirds of usage contexts employ forwarder proxy contracts, whereas 32.2\% that utilize upgradeability proxy contracts.}\label{obs:rq3-01} Out of 385 representative proxies, 124 (32.2\%) were designed for upgradeability purposes, whereas 261 (67.8\%) proxies only forward calls to a logic contract. As mentioned earlier in the introduction, monitoring proxies and especially upgradeability proxies is critical to the security of many applications. As such, our findings highlight the need for tools that monitor upgradeability proxies for the replacements of logic contracts and the proxy admin. Monitoring changes in the proxy admin address is important for governance because it ensures that only authorized parties have the ability to make changes or upgrades to the code \citep{eip1967}

\begin{figure}[!t]
    \centering
    \includegraphics[width=0.85\columnwidth,]{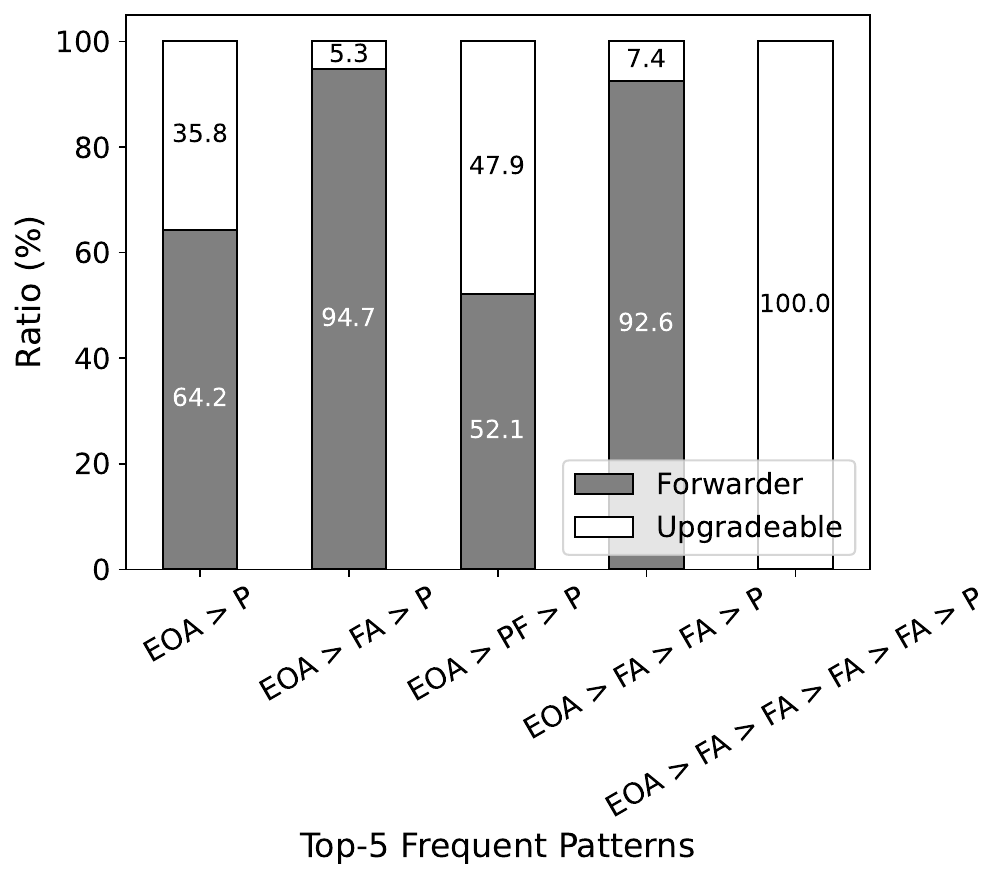}
    \caption{The ratio of different proxy types across the top-5 patterns in terms of the number of deployed proxies.}
   \label{fig:rq3-01}
\end{figure}

\smallskip \observation{Upgradeability and forwarder proxies are deployed through both on-chain and off-chain styles.}\label{obs:rq3-02} Figure \ref{fig:rq3-01} compares the ratio of usage contexts with either upgradeability and forwarder proxies across the top-5 creational patterns. It is evident that the majority of usage contexts use forwarders for the top-4 patterns with the exception of the fifth pattern whose all usage contexts employ upgradeability proxy contracts. Concerning upgradeability proxies, we found instances of this class across the top-5 patterns, albeit to different extents. More specifically, two on-chain patterns (i.e., \textit{EOA > FA > P} and \textit{EOA > FA > FA > FA > FA> P}) have a higher ratio of upgradeable usage contexts (47.9\% and 100\%, respectively) compared to the only off-chain pattern (i.e., \textit{EOA > P} with 35.8\%). On the other hand, there are two other on-chain patterns (i.e., \textit{EOA > FA > P} and \textit{EOA > FA > FA > P} with 5.3\% and 7.4\% in turn) with mostly usage contexts that employ forwarder proxy contracts. Finally, we perform a chi-squared test ($\alpha=0.05$) with a null hypothesis that there is \textit{no association} between the creational patterns and the proxy types. Corroborating our visual analysis of Figure \ref{fig:rq3-01}, the test result indicates that we can reject the null hypothesis (p-value $\leq$ 0.05).

\smallskip \observation{Just below 30\% of usage contexts employed ERC-1167 minimal proxies, while around 21\% used customized implementations.}\label{obs:rq3-03}As per the results yielded by the evm-proxy-identification tool, the majority of usage contexts (79\%) adhere to the best practices while adopting proxy contracts. More specifically, the ERC-1167 minimal proxy implementation dominates with a ratio of approximately 29.38\%, closely followed by ERC-897 and ERC-1967, representing about 25.08\% and 22.91\%, respectively. EIP-1167 has had a great impact on reducing deployment costs when cloning contract functionality \citep{eip_1167}. Thus, we conjecture that the majority of proxies in Ethereum are for reducing deployment costs. Surprisingly, a significant portion, around  21.1\%, were classified as ``Customized'', indicating that these proxy contracts did not adhere to any of the recognized implementations (see Section \ref{sec:threats} for a discussion of possible reasons). The least adopted implementations were the ERC-1967 (Beacon), ERC-1822, and the Gnosis Safe Proxy with ratios of 0.28\%, 0.25\%, and 0.13\%, respectively.

\begin{figure}[!t]
    \centering
    \includegraphics[width=0.85\columnwidth,]{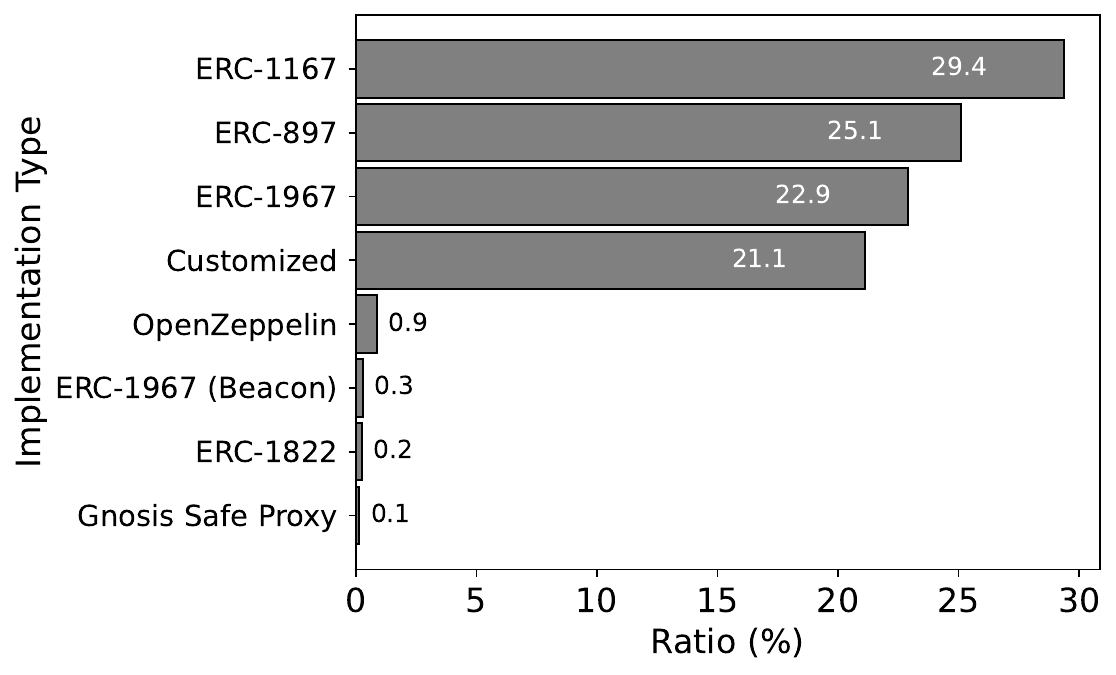}
    \caption{The ratio of different proxy reference implementations.}
   \label{fig:rq3-02}
\end{figure}

\begin{footnotesize}
    \begin{mybox}{Summary}
        \textbf{RQ3: \RQThree}
        \tcblower
        a predominant use of forwarder proxies (67.8\%) and a significant presence of upgradeability proxies (32.2\%), highlighting the importance of monitoring tools for security and governance. The deployment of these proxies spans both on-chain and off-chain styles, with certain patterns favoring upgradeable proxies. Additionally, while most proxies adhere to best practices like ERC-1167, a notable 21.10\% are customized, indicating a diverse range of implementations. 
    \end{mybox}
\end{footnotesize}  
    \section{Discussion}
\label{sec:discussion}

\subsection{Advantages of using the proxy pattern in DApps}\label{subsec:benefits-of-the-proxy-pattern}
\begin{itemize}[label=\textbullet, itemsep = 3pt, topsep = 0pt]
    \item \textbf{Enhanced modularity and maintenance}. The use of proxy contracts in DApps enhances their modularity and maintenance capabilities in a seamless and efficient manner. By separating logic from data, these contracts ensure a clear architectural division that not only increases flexibility but also simplifies modifications and upgrades. This is exemplified in the ``Diamond Standard'' or EIP-2535, which stands as a testament to the advanced modularity achievable through this pattern. Simultaneously, the Proxy pattern facilitates the easy swapping of logic contracts. This adaptability allows developers to continuously update and refine a running DApp, introducing new features and addressing bugs, thereby transforming it into a more maintainable and agile software solution.
    
    \item \textbf{Reducing deployment cost}. A proxy contract can reduce deployment gas costs by centralizing the main logic in one contract (the implementation contract) and deploying lightweight proxy contracts (e.g., a minimal proxy contract) for each instance. These proxy contracts simply delegate calls to the main logic contract. Since each proxy contract only contains the delegation mechanism and not the replicated logic, the gas cost for deploying these proxies is significantly lower than that for deploying the full logic contract multiple times. See Appendix \ref{apx:reducing-dep-cost-with-minimal proxy-contracts} for a case study of reducing deployment gas costs using proxy contracts.  
    
    \item \textbf{Third party contract reuse}. The proxy pattern allows developers to reuse an already deployed contract (e.g., the Gnosissafe wallet contract\footnote{\url{https://etherscan.io/address/0xd9db270c1b5e3bd161e8c8503c55ceabee709552}}) by setting up a proxy contract that forwards calls via \texttt{delegatecall} to that target contract. This method enables the proxy to mimic the target contract's behavior by executing its logic within the proxy's context. It is vital to align the storage layouts and be mindful of the original contract's permissions, which could restrict interactions.
    
    \item \textbf{Continuity}. One of the primary benefits of using proxy patterns is the ability to upgrade or modify the proxy’s logic without requiring the deployment of a new contract address. This means that users can interact with the same proxy contract address consistently, ensuring a seamless experience even if the underlying logic changes. This creates a more predictable and stable environment for users.
    
    \item \textbf{Data persistence}. Traditional contract upgrades without a proxy pattern often require deploying an entirely new contract, migrating data, and instructing users to interact with a new address. However, with a proxy pattern, only the logic contract is upgraded, and the data remains intact in the proxy contract. This can be much more gas-efficient than deploying a new contract and migrating data.

    \item \textbf{Limited exposure}. The proxy pattern's division of a contract system into distinct parts adds a layer of defense. This separation means that if a vulnerability arises in the logic, it may lead to unintended behavior without directly compromising the stored data or assets. Conversely, a flaw in the storage might put data at risk, but not necessarily enable the execution of malicious functions from the logic. Additionally, by having these multiple layers, the system gains an extra security buffer. If one layer is compromised, the other can still provide protection. A proxy can be configured to allow only certain types of calls to the logic contract, enhancing overall security through this layered approach.
\end{itemize}

\subsection{Drawbacks of proxy contracts in DApps}\label{subsec:drawbacks-of-the-proxy-pattern}
\begin{itemize}[label=\textbullet, itemsep = 3pt, topsep = 0pt]
    \item  \textbf{Architectural complexity}. The proxy pattern introduces an additional layer to DApps which can potentially increase the transaction processing time. In addition, Instead of a single contract, developers now have to maintain at least two contracts: the proxy and the logic contract. This can lead to higher maintenance burdens, as developers need to monitor, update, and ensure the compatibility of both the proxy and logic contracts across different versions.
    \item  \textbf{Gas overhead}. Although proxy contracts can offer gas optimizations in some cases (e.g., ERC-1167 minimal proxies or data migration), the additional indirection can also introduce gas overheads for certain operations due to the extra delegation call. In Solidity, the \texttt{delegatecall} operation carries a base gas cost of 700. Additionally, transmitting data with the \texttt{delegatecall} incurs variable gas costs.
    \item  \textbf{Centralization \& governance concerns}. Centralization and governance challenges arise in proxy contracts if upgradeability is not properly implemented. A single entity with upgrade control could introduce potential vulnerabilities or malicious changes. Furthermore, for decentralized systems, achieving consensus on upgrades can be complex and prolonged, adding to trust and operational concerns.
    \item  \textbf{New attack vector}. While proxy contracts can help in isolating and addressing vulnerabilities, they also introduce new potential attack vectors. For instance, if an attacker gains control of the proxy's admin privileges, they can redirect the proxy to any logic they want. Several incidents have been reported regarding proxy contracts \citep{attack_vector_proxy}. For instance, a breach into the admin's private key, as seen in the PAID Network incident, can result in significant money losses. In that specific case, an attacker compromised the proxy admin's private key and modified the logic contract, allowing them to burn and mint PAID tokens at will.
    \item \textbf{Transparency concerns}. For end-users, it might not always be clear that they are interacting with a proxy contract, which can lead to confusion or misinterpretation of the contract's actual functionality, especially if the logic is changed without the user's knowledge. We will further discuss this issue in Section \ref{subsec:discussion:implication}.
\end{itemize}

\subsection{Benefits and drawbacks of on-chain deployment style}\label{subsec:benefits-and-drawbacks-of-on-chain}
We term a pattern ``on-chain'' when a smart contract, known as the factory, creates instances or clones from another smart contract template during runtime. The contract template's bytecode can be stored off-chain and sent to the factory during runtime or embedded directly within the factory. To safeguard templates and maintain consistency in the clones, we recommend the embedded approach, especially for generating multiple instances. An on-chain factory smart contract offers several benefits, both from the perspectives of developers and users:
\begin{itemize}[label=\textbullet, itemsep = 3pt, topsep = 0pt]
    \item \textbf{Gas efficiency with clone factories}. Clone factories can employ the EIP-1167 minimal proxy pattern to minimize deployment gas costs. A master contract is deployed only once, and the factory spawns a typically smaller and cheaper proxy contract per request/user. This process allows end-users to reuse the master contract's functionality while having control over their own contract's storage and state. GnosisSafe (Appendix \ref{apx:discussion-of-largest-dapps-for-the-top-5-patterns}) and Uniswap-V1 (Appendix \ref{apx:reducing-dep-cost-with-minimal proxy-contracts}) are two DApps that use clone factories to reduce deployment costs.
    \item\textbf{Standardization}. Clone factories ensure that every contract spawned follows a standardized template, promoting consistency in behavior and ensuring users can trust the contract's operations. 
    \item\textbf{Reusability}. Some factories can receive bytecode for any intended contract and deploy it, which promotes reusability and reduces deployment efforts required for developing off-chain deployment scripts. Such factories advocate for reusability and reduce the deployment efforts required for developing off-chain deployment scripts. The longer chains in Table \ref{tab:rq2-01} can be the result of reusing such general-purpose factories.
    \item\textbf{Interoperability}. Factory contracts can be designed to integrate or communicate with other contracts or systems on the blockchain, enabling seamless interoperability and richer functionalities.
    \item\textbf{Transparency \& trustworthiness}. The entire process, from the contract creation to its interactions, is recorded on the blockchain. This transparency builds trust among users, as they can independently verify the code and the instantiation process of the contract. 
    \item\textbf{Simplified management}. Developers can manage and update the factory contract. If there is a need to make upgrades or modifications, the developer can address the factory contract, which can then influence all subsequent contracts it spawns. We detected the PlotX DApp where developers employed several layers of factory contracts to manage DApp’s contracts at different levels (Appendix \ref{apx:discussion-of-largest-dapps-for-the-top-5-patterns}).
    \item\textbf{Enhanced security}. On-chain factories can be designed to incorporate security checks, ensuring that the spawned contracts adhere to certain safety standards or practices. For instance, a factory can have different creation methods that deploy instances under different circumstances. 
\end{itemize}

\smallskip While on-chain factory smart contracts bring several benefits, they also come with potential drawbacks and challenges:
\begin{itemize}[label=\textbullet, itemsep = 3pt, topsep = 0pt]
    \item \textbf{Initial overhead}. Setting up a factory smart contract can have higher initial costs (in terms of development effort and gas fees) compared to deploying a standalone contract.
    \item \textbf{Architectural complexity}. Developing and managing a factory contract can be more complex than a singular, standalone contract. This complexity might pose challenges for developers unfamiliar with the factory pattern.
    \item \textbf{Single point of failure}. If there is an error or vulnerability in the factory, every contract it spawns might inherit that issue. Addressing this can be challenging, especially if many contracts have already been deployed.
    \item \textbf{Centralization concerns}. If not designed carefully, the factory pattern might introduce centralization. For instance, if the factory has an admin with elevated privileges, it could become a central point of control or failure.
\end{itemize}

\subsection{Benefits and drawbacks of off-chain deployment style}\label{subsec:benefits-and-drawbacks-of-off-chain}
In contrast to the on-chain deployment method where proxies are deployed by a factory contract, the off-chain style deploys them directly using an EOA and a deployment script. The key distinction between the two methods is the location of the deployment infrastructure. On-chain relies on a decentralized factory contract, whereas off-chain uses centralized, locally-maintained scripts, like those in the Truffle framework. When needed, the proxy is then deployed to Ethereum by the EOA via running the deployment scripts. Off-chain deployment style offers several advantages, including:
\begin{itemize}[label=\textbullet, itemsep = 3pt, topsep = 0pt]
    \item \textbf{Cost efficiency}. By handling heavy computations off-chain, one can save on gas costs associated with on-chain computations. When scaling to a large number of deployments with a computationally-intensive initialization process, this can result in significant savings.
    \item \textbf{Flexibility}. Deploying from off-chain scripts provides developers with a greater degree of control and adaptability. They can modify deployment parameters, test in various environments (like Ethereum testnets), and easily customize the deployment process without being constrained by the blockchain's rules and limitations.
    \item \textbf{Scalability}. Off-chain style can handle large-scale deployments and interactions by leveraging off-chain infrastructure and services, whereas on-chain solutions might be constrained by the blockchain's processing power and gas limits. For instance, off-chain deployment scripts can be set up to deploy multiple contracts in parallel, leveraging powerful centralized servers or cloud infrastructure. This is in contrast to on-chain deployments that might be restricted by blockchain's inherent sequential processing.
    \item \textbf{Batch deployment}. Off-chain solutions can accumulate multiple deployment requests and batch them into fewer on-chain transactions. This way, multiple operations can be processed in one go, optimizing the use of network resources.
    \item \textbf{Integration with other development tools}. Off-chain scripts, especially those written for frameworks like Truffle, can be seamlessly integrated with other developer tools, version control systems, and continuous integration/continuous deployment (CI/CD) pipelines.

\end{itemize}

\smallskip While deploying contracts using the off-chain style has several advantages, it also comes with certain drawbacks:

\begin{itemize}[label=\textbullet, itemsep = 3pt, topsep = 0pt]
    \item \textbf{Lack of transparency}. Unlike on-chain deployment methods, off-chain deployment does not record every preliminary action on the blockchain. As a result, the entire deployment process might not be as transparent to external observers.
    \item \textbf{Centralization risks}. Off-chain scripts are typically maintained and executed by a centralized entity or group. This central point of control could become a target for attacks or misuse.
    \item \textbf{Trust issues}. Since the deployment is managed off-chain, other parties must trust that the deploying entity has not made any malicious or erroneous modifications before deploying the contract on-chain.
    \item \textbf{Security concerns}. Deployment scripts, if not securely managed, could be tampered with or compromised. This can lead to the deployment of malicious contracts or the loss of valuable assets.
\end{itemize}

\subsection{Implications to practice} \label{subsec:discussion:implication}

\implication{Research should gradually shift towards a system view while analyzing smart contracts.} Our motivation study (Section \ref{Sec:prem}) has revealed that smart contracts are becoming increasingly interdependent, with many contracts being highly coupled to other contracts. This highlights a paradigm shift in smart contracts' design: from monolithic contracts to modular systems. Therefore, it is imperative for researchers to consider a system view perspective (i.e., the set of contracts that belong to a larger context or DApp), rather than focusing solely on individual contracts. However, to date, no systematic research has been conducted to determine the scope of systems in Ethereum. Thus, future studies should aim to explore the systems' scope in more detail. 

As for the proxy contracts, they add a layer of indirection, and we showed that such contracts have been playing a crucial role in smart contract interactions (Observation \ref{obs:rq1-04}) since the early days of Ethereum. In our study, we introduced the ``usage context'' notion as a way of grouping proxy contracts into clusters, each representing a set of similar proxy contracts that are deployed by an identical entity and whose purposes are similar. We then use this concept while analyzing proxy contracts throughout our study. We brought some examples of large decentralized systems that use the top-5 on-chain proxy creational patterns and discuss their purpose (Appendix \ref{apx:discussion-of-largest-dapps-for-the-top-5-patterns}). In addition, we analyzed the application of proxy contracts employed by the top-50 most active DApps (Appendix \ref{subsec:application-of-proxy-pattern-in-dapps}). However, there is a need for additional research to comprehensively explore the characteristics of the systems supporting the entirety of the 7.2 million detected proxies.

\smallskip \implication{Proxy contracts reduce transparency in DApps.} Proxy contracts can reduce transparency in DApps by adding an extra layer of indirection between users and the underlying logic contract. While this can enhance modularity, it may decrease transparency if essential information about the logic contract (e.g., source code or address) is not accessible. Having studied the practical applications of proxy contracts in DApps (Appendix \ref{subsec:application-of-proxy-pattern-in-dapps}), we found that out of 307 DApps with at least one proxy contract, 297 did not publish some or any of their logic contract addresses. Specifically, 97.1\% of proxies lacked published logic contract addresses. Several reasons could explain this phenomenon. Developers might aim to protect against potential attacks on the logic contract or safeguard their intellectual property. Some may not fully grasp the importance of sharing logic contract addresses, while others using upgradeability proxies might avoid publication due to frequent changes. Future research could delve deeper into developers' motivations. Similarly, we found that over 20\% of usage contexts use proxy contracts that do not follow the storage layout of popular proxy reference implementations (Section \ref{sec:rq3}). This can prevent blockchain explorers (e.g., Etherscan) from tracking the logic contract of such contracts, further contributing to this issue. 

This finding underscores a few critical points. Transparency is vital for DApps, and the absence of logic contract addresses complicates auditing. Additionally, for developers seeking to interact with the logic contract via its proxy, identifying the right functions to call becomes challenging, hindering reusability. Lastly, incomplete contract information in marketplaces poses challenges for blockchain researchers in system view analysis, as these marketplaces are a primary source for DApps' contract lists.

Given that proxy contracts can inadvertently reduce transparency, it is imperative for DApp developers to prioritize the publication of logic contract addresses alongside proxy addresses, allowing users, auditors, and fellow developers to access vital information about the core functionality and interfaces of your DApp. In addition, marketplaces can utilize services like Etherscan Proxy Verification to display a proxy's current logic contract. Finally, our proxy detection method (Section \ref{sec:proxy-detection}) can help marketplaces identify proxies and their logic contracts for automatic updates of DApps' missing contracts.

\smallskip\implication{Developers should weigh the benefits and challenges of the proxy pattern and choose the deployment styles that best meet the needs of their DApps.} When deciding to employ the proxy pattern, developers face a nuanced trade-off between several factors. On the one hand, the proxy pattern offers substantial benefits like flexible upgrades, cost reductions in large-scale deployments etc., as outlined in Section \ref{subsec:benefits-of-the-proxy-pattern}. On the other hand, it also introduces architectural complexity and the need for rigorous design, consistent audits, and a transparent governance structure to mitigate potential risks (Section \ref{subsec:drawbacks-of-the-proxy-pattern}). Key considerations in this decision include the DApp's characteristics, anticipated upgrade frequency, user base size and characteristics, and the value of user trust. For DApps where long-term maintenance, data longevity, modularity, or gas-saving deployment at scale are essential, the proxy pattern might be highly suitable. Conversely, simpler DApps or those for which transparency is of utmost importance may find the added complexity of the proxy pattern unnecessary.


When choosing between on-chain and off-chain styles for deploying proxy contracts, developers are presented with a nuanced trade-off. On one hand, on-chain deployment, with its factory contracts, offers benefits such as gas efficiency, standardization, reusability, interoperability, transparency, simplified management, and enhanced security (Section \ref{subsec:benefits-and-drawbacks-of-on-chain}). It is particularly advantageous in larger usage contexts where standardization and consistent deployment patterns are essential. On the other hand, off-chain deployment provides cost efficiency in handling heavy computations, increased flexibility, scalability, batch deployment capabilities, seamless integration with other development tools, and enhanced privacy (Section \ref{subsec:benefits-and-drawbacks-of-off-chain}). It lends itself well to scenarios where customized deployments, rapid iterations, or integration with broader software ecosystems are essential. However, it comes with its challenges, such as potential centralization risks and decreased transparency. Thus, the decision largely hinges on the specific requirements and constraints of a DApp, with developers needing to weigh the immediate advantages of a chosen style against its long-term implications and potential challenges.

Based on the provided analysis (Observation~\ref{obs:rq2-dep-style-and-contet-size}), developers appear to lean towards on-chain deployment when dealing with larger usage contexts. The one-time gas expense of deploying factories, despite its initial overhead, proves to be more cost-effective in scenarios with multiple proxies (Appendix~\ref{subsec:gas-cost-complexity-of-dep-style}). Conversely, for smaller usage contexts, especially singletons, off-chain deployment methods are favored. These off-chain methods, although having a larger bytecode size, hint at a richer functionality, likely because developers are less concerned with gas optimization in these scenarios and more inclined to offer comprehensive features. However, as the size of the usage context grows, the trend seems to favor more gas-efficient deployments, perhaps using minimal proxy contracts to reduce gas costs. While the data presents clear patterns in developers' deployment preferences based on context size and gas costs, it is paramount for future research to delve deeper into how the total deployment gas cost, complexity of initialization, and storage costs impact the choice of deployment style and to ascertain the magnitude of the effects of such variables.

\smallskip \implication{Future studies should delve into the release engineering process of smart contracts through proxies.} \citet{Chen20-2} explored the reasons why developers terminate smart contracts. Their method, rooted in code similarity, first detects a self-destructed contract before identifying its subsequent version. The \texttt{SelfDestruct} feature in Solidity lets a contract be destroyed, transferring its balance to a specified address. Improper use can lead to security vulnerabilities or funds loss. While \citet{Chen20-2}'s study pioneered understanding releases, it did not address release engineering via proxy contracts. \citep{wohrer2021devops} studied how DevOps can be applied to blockchain-based software development, especially for Ethereum smart contracts. The authors proposed a structured DevOps procedure that covers the main stages of Continuous Integration and Continuous Delivery along with a discussion around the challenges and benefits of DevOps for blockchain, such as the need for more rigorous testing and differentiated deployment due to the immutability of blockchain. While this study focused on core DevOps practices, it is worth noting that there remains a gap in our understanding of the characteristics of releases through the proxy pattern method.

Proxies, especially those that are upgradeable, are increasingly favored for their adaptability in crafting update-friendly contracts. From our manual investigation, over 32\% of usage contexts contain upgradeability proxy contracts (Observation \ref{obs:rq3-01}). However, a deeper dive is required to fully grasp release characteristics. To jumpstart future research on upgradeable contracts, we suggest six pivotal questions:


\begin{enumerate}[label=\textbullet{ }Q\arabic*., itemsep = 3pt, topsep = 0pt, wide = 0pt, font=\itshape]
    \item How prevalent are different proxy types? Even though we conducted a manual study on a representative sample of proxy contracts (Section \ref{sec:rq3}), there has been no large-scale study that automatically analyzes the prevalence of upgradeable vs forwarder proxies since Ethereum's birth. In addition, we could not detect the reference implementation of around 21\% of proxy contracts using evm-proxy-detection tool. In light of this observation, how static analysis techniques can be specifically tailored to detect these variations effectively?
    
    \item In light of our first implication, how many DApps use upgradeable vs forwarder proxies? 
    
    \item How is the distribution of the number of releases for upgradeable DApps? The number of times a new version of a logic contract is replaced an old version can indicate the extent to which release engineering is a common practice in the Ethereum ecosystem. 
    
    \item What is the average amount of time required to upgrade a DApp?
    
    \item What techniques are used to ensure upgradeable governance in DApps? Smart contract upgrades violate immutability and must be conducted in a secure, controlled, and transparent manner. Therefore, it is important to establish upgradeable governance. Upgradeable governance refers to the rules and processes for managing and executing smart contract upgrades. Several governance mechanisms exist (e.g., on/off-chain voting, multi-sig contract, EOA, etc.) \citep{Liu23,openzeppelin_govern}. Thus, it is important to analyze how developers govern contracts (and hence DApp) upgrades. In other words, how do developers maintain the trust of their end-users when they practice upgradeability?
    
    \item To what extent is a suitable governance technique used that aligns with the size of a DApps' user base? The higher the number of users, the more critical and sensitive a contract upgrade could be. For instance, for a given DApp with thousands of users, if the upgrade is controlled by a single EOA, then it is a sign of bad governance practice. It is also interesting to analyze how the size of the user base affects the choice of upgradeability governance in existing DApps. Therefore, it is important to study the employed governance technique while taking the contracts' user base into account.
    
\end{enumerate}

Addressing these queries demands automated tools for spotting upgradeable proxies and determining governance methods. Some research has begun in this direction~\citep{Salehi22}, yet there is ample room to further explore the breadth of release engineering in Ethereum.

    \section{Threats to Validity}
\label{sec:threats}

\noindent \textbf{Construct Validity.} Detecting proxies is not a trivial task. Similar to \citet{Salehi22}, we proposed an approach that detects proxies based on their behavior. Since \citet{Salehi22} did not provide a replication package, we re-implemented their approach. While we were not able to thoroughly compare our approach with the previous study, we tested and evaluated it to the best of our abilities to ensure its correctness (Section \ref{sec:evaluation}). We acknowledge that, while a behavioral detection approach has high performance, it cannot detect inactive proxies (i.e., proxy contracts whose proxy functionality is never used). Additionally, we used the Panoramix tool to decompile the bytecode of the contracts during our evaluation. This tool has been utilized by Etherscan \footnote{\url{https://etherscan.io/bytecode-decompiler}}, the most prominent blockchain explorer, as well as in prior research \citep{Salehi22}. While our focus is on active proxy contracts, we used a ground truth dataset of both active and inactive proxy contracts to evaluate our proxy detection method. The reasons are twofold. Firstly, even though our focus is on the active proxies, we intended to shed light on the actual ratio of proxy contracts by creating a ground truth that included both active and inactive proxy contracts. As such, we found that 23.3\% of our sample contracts were indeed proxies. Given we found 14\% of contracts are active proxies (Observation \ref{obs:rq1-01}), this means that just above 9\% of proxies are inactive. Secondly, future research can fruitfully use our ground truth dataset for evaluation regardless of the characteristics of their detection method. 

In RQ3, we relied on a tool called \textit{evm-proxy-detection} to distinguish between seven proxy reference implementations (Section \ref{sec:rq3}). This tool is developed by Gnosis\footnote{\url{https://www.gnosis.io/}}, which is a reputable open-source, decentralized prediction market built on the Ethereum blockchain. Different reference implementations have a unique signature (e.g., either their storage structure or bytecode) that the tool relies on in order to detect the proxy implementation type. Given that the tool detects the type of the majority (79\%) of our proxies, we believe that it still provides useful insight into the prevalence of different reference implementations. While the tool could not detect the type of 21\% of representative proxies, we acknowledge that this does not mean that those proxies are poorly implemented. Indeed, although many of the reference implementations emerged after 2018, Figure \ref{fig:rq1-usage-context} shows that proxies have emerged since mid-2016 (Observation \ref{obs:rq1-02}). Thus, it is natural that not all proxy contracts are based on known reference implementations. Upon analyzing a subset of these ``Customized'' cases, we found that slight bytecode variations from reference implementations hinder detection by the tool, especially for minimal proxy contracts. Additionally, some developers opt for custom designs, deviating from standard storage layouts, and making their contract type undetectable by the employed tool.


\smallskip \noindent \textbf{Internal Validity.} To mitigate internal validity threats, we explicitly consider code cloning in Ethereum while studying the prevalence of proxy contracts in the first research question. We also mitigate other internal validity threats by ensuring that our chosen statistical analysis procedures properly suit our goals and the characteristics of our datasets. In addition, in Section~\ref{sec:discussion}, we collected DApps' contract list from three different marketplaces to reduce the risk of missing data points.

Furthermore, in our manual classification proxies based on their purpose (RQ3), we followed open coding best practices and relied on an approach proposed by a previous study \citep{Salehi22}. From a statistical point of view, we highlight that we randomly picked a statistically representative sample of proxies in our analyses. In addition, prior to classifying our proxy instances, we carefully reviewed different proxy types in the literature~\citep{eip_897,eip_1167,eip1967,openzeppelin_proxy,eip_1822}. Such a review informed and supported the entire coding process. Finally, we also leveraged our own expertise in smart contracts while performing the classification~\citep{Oliva20a, Oliva20b, Oliva20c, Oliva21, Oliva23a, Oliva23b}. 

\smallskip \noindent \textbf{External Validity.} The dataset used in this work is a collection of \textit{all} smart contracts and their transactions that were available on Etherscan as of September 1st, 2022. As such, our dataset is a complete representation of contracts at that point in time. Nevertheless, it is possible that smart contracts developed after such a period may have different characteristics. It is thus possible that our results do not generalize to newer contracts. Also, our results might not generalize to other programmable blockchain platforms. Finally, we only focus on active proxies. However, it is possible that inactive proxies may have unique characteristics. Therefore, our results may not generalize to inactive proxy contracts.

    \section{Related Work} 
\label{sec:related-work}

\smallskip \noindent \textbf{Proxy pattern in other disciplines.} Design patterns provide standardized solutions to recurring software design issues, facilitating an efficient response to common challenges in system creation. The prominence of design patterns in computer science grew substantially after the Gang of Four (GoF) published a book on standard patterns for object-oriented systems~\citep{gof_book95}. Subsequent research has expanded on these concepts, reusing existing patterns or identifying new ones across various software domains, including microservices architectures~\citep{Richardson18}, IoT systems~\citep{Bloom18,Ngaogate19}, and cloud computing~\citep{indrasiri2021design}.

Proxies, as design structures, serve varied roles across different system contexts. In object-oriented systems, they can mask the intricacy of primary objects, facilitating functionalities such as lazy initialization, access control, and logging. 
In distributed systems, proxies are often advised to add structure and encapsulation to the system~\citep{Shapiro86}. For instance, in microservices, they form the core of the API gateway\footnote{\url{https://microservices.io/patterns/apigateway.html}} pattern, which acts as the external entry point into applications. Thus, this underscores the proxy's indispensable role in both conventional and contemporary software paradigms.

\smallskip \noindent \textbf{Proxy pattern in smart contract design.}  \citet{Rajasekar20} explored 19 distinct design patterns for blockchain applications, categorized into five domains: security, data, creational, structural, and behavioral. They emphasized the Proxy pattern, which facilitates contract modification and updates, and the Factory contract, aiding contract creation on the blockchain.\citet{Wohrer18} identified 18 smart contract design patterns through a Multivocal Literature Research and a qualitative analysis, categorizing them into five groups. They discussed the proxy mechanism within the relay contract pattern, enabling contract upgrades.\citet{Xu21} presented several design patterns along with their merits, demerits, and use cases. They proposed a decision model to assist in selecting appropriate patterns for decentralized applications, discussing proxy patterns and factory contracts for contract maintenance and on-chain contract generation.\citet{Kannengiesser21} conducted a study identifying 29 smart contract development challenges and 60 solutions, extracting 20 design patterns, including detailed discussions on proxy and factory patterns.\citet{Worley19} proposed design patterns to address common smart contract constraints, discussing the proxy pattern within the migration patterns.

\smallskip \noindent \textbf{Proxy contract detection.} Detecting proxies, although challenging, is crucial for many applications' security~\citep{eip_897}. The most related study is by~\citet{Salehi22}, introducing a method to detect proxy contracts based on run-time behavior. Their analysis spanned from Sep-05-2020 to Jul-20-2021. In contrast, our study covers Ethereum's entire lifespan, from Aug-01-2015 to Sep-01-2022. \citet{Salehi22} utilized an Ethereum full archival node to replay transactions and extract traces. This method, while accurate, is resource and time-intensive. Instead, we harnessed the Ethereum public dataset on Google BigQuery, proposing a more efficient technique that identifies all proxy contracts in under 15 minutes without replaying transactions. A comparison with \citeauthor{Salehi22}'s method is available in Section \ref{sec:evaluation}. Additionally, while~\citet{Salehi22} initiated an automated method to discern upgradeability and forwarder proxies, the limited analysis period may affect their generalizability. We address this by meticulously analyzing a statistically representative proxy sample on Ethereum (Section~\ref{sec:rq3}), enhancing the robustness of our findings.

Etherscan\footnote{\url{https://etherscan.io}}, the most prominent blockchain explorer, offers an online service called ``Proxy Verification''\footnote{\url{https://etherscan.io/proxyContractChecker}} that determines whether a deployed contract is a proxy or not. We did not use this service due to two reasons. First, one can only send 100 requests per day to the specified end-point. As such, it is impossible to use it for such a large-scale study. Second, as acknowledged by Etherscan, this service is based on a heuristic. Since the details of the heuristic are not available online, we could not rely on it. In contrast, our proposed approach identifies proxies through two definitive features of a proxy contract. (Section \ref{sec:background}).

Finally, we also found a tool called \textit{Evm-Proxy-Identification} \citep{evm_proxy_detection}. We use this tool to categorize different proxy implementations in Section \ref{sec:rq3}. Yet, we could not use this tool for detecting proxies in general due to several reasons. First, the tool uses a blockchain API called Infura\footnote{\url{https://www.infura.io/}}, which allows a limited number of requests per day for a free account. Second, this tool detects certain popular proxy types. For instance, as discussed in Observation~\ref{obs:rq3-03}, over 21\% of our detected proxies are labeled as ``Customized'' by the tool, which indicates the limited capability of the tool in detecting proxies. In contrast, our approach detects proxies regardless of their type by relying on their behavior, which is similar across all proxy types, implementations, and standards.

We also examined the \textit{Evm-Proxy-Identification} tool \citep{evm_proxy_detection} but faced limitations due to its reliance on the restricted Infura API and its constrained proxy detection. Our method, conversely, identifies proxies across types and implementations by focusing on consistent behavior.

    \section{Conclusion}
\label{sec:conclusion}


The proxy pattern is a well-established design pattern in software design with numerous use cases in various disciplines. In the context of programmable blockchain, the role of proxies in facilitating smart contract maintenance and reducing contract deployment is undeniable. Yet, little is known about the characteristics of the proxy pattern in this context.


In this paper, we conducted a large-scale study of the proxy pattern in a dataset of 50,845,833 smart contracts. As the first large-scale study in the area, we attempted to characterize the proxy pattern by determining its prevalence (RQ1), different ways of its integration (RQ2), and its different types (RQ3). Using a behavioral detection technique, we found that over 14\% of the contracts are active proxies. 

In RQ1, We observe an upward trend in the usage of proxy contracts over the entire Ethereum lifespan from various perspectives, with its notable role in increasing modularity. 

In RQ2, we found 12 different creational patterns for deploying proxies, which we classified into \textit{off-chain} and \textit{on-chain} styles depending on where the deployment script is operating. We realized that on-chain patterns are the most popular way of creating proxies in different contexts (i.e., proxy contracts of around 61\% of usage contexts are deployed through this style). Yet, the most popular off-chain pattern is as popular (39.07\%) as the most popular pattern of on-chain deployment style (39.81\%).

In RQ3, we found that while the majority of proxies (67.8\%) typically intercept calls, process, and subsequently forward them to their logic, 32.2\% enable contract maintenance and upgrades. Both forwarder and upgradeable are deployed through both on-chain and off-chain patterns. Finally, we found that the majority (79\%) of usage contexts employ proxies that are implemented based on the best practices and standards. In particular, just below 30\% of the usage contexts use ERC-1167 minimal proxies, which is beneficial for reducing deployment costs.

Furthermore, we thoroughly discussed the benefits and drawbacks of the proxy pattern, off-chain, and on-chain deployment styles (Section \ref{sec:discussion}). In addition, we studied the practical applications of the proxy pattern in the top-50 most active DApps and found that the majority of them use proxy contracts for upgradeability purposes (Appendix \ref{subsec:application-of-proxy-pattern-in-dapps}). We also revealed that in spite of the importance of transparency for decentralized applications, DApps often do not publish the logic contract address to which their proxies delegate calls.

Finally, based on our findings, we discussed opportunities for future research and implications to practice (Section \ref{sec:discussion}). In particular, we (i) highlighted the lack of techniques for understanding smart contracts at the system level, (ii) proxy contracts decrease transparency in marketplaces, (iii) shed light on the usage of on-chain and off-chain styles along with their trade-offs, and (iv) provided a research agenda with six open questions for further studies. We hope that our study will inspire and bootstrap additional studies regarding the usage of proxy contracts, especially in the field of release engineering for DApps.

    \section*{Data Availability Statement (DAS)}
\label{sec:Data_Availability_Statement}


A supplementary material package is provided online\footnote{\url{https://github.com/SAILResearch/replication-24-amir-smart_contracts_proxy-code-emse}}. The contents will be made available on a public GitHub repository once the paper is accepted.

    \section*{Conflict of Interest (COI)}
\label{sec:Conflict_Of_Interest}
The authors declared that they have no conflict of interest.
    
    \bibliographystyle{IEEEtranN}
    
    \bibliography{10-references}
    
    \clearpage 
    \begin{appendices}
        \section{Practical applications of the proxy pattern in DApps}\label{subsec:application-of-proxy-pattern-in-dapps}
To gain a more holistic understanding of the usage of proxy patterns, we investigated those in the context of DApps. More specifically, we analyzed the contract list of 3,767 Ethereum DApps collected from three well-known marketplaces, namely stateofthedapps.com, dappradar.com, and dapp.com. Having cross-referenced the DApps' contracts with our proxy contract dataset (Section \ref{sec:rq1}), we found that 307 (8.14\%) DApps have at least one proxy contract. We then sorted these 307 DApps based on their activity level in ascending order. To determine the activity level of a given DApp, we calculated the average number of inbound transactions received by all of the DApps' contracts as of September 5th, 2023. First, we determined the date when the very first contract of a DApp was deployed, which we refer to as the DApp's 'birthday'. Then, we computed the DApp's age by counting the number of days from its birthday to September 5th, 2022. Subsequently, we tallied the number of inbound transactions that all the DApps' contracts received during this period, which we refer to as the 'inbound traffic'. Finally, we divided a DApp’s inbound traffic by its age to derive the activity level. 

To understand how proxies are used in relevant DApps, we selected the top-50 most active DApps. Subsequently, we analyzed the proxy contracts of these DApps (a total of 2,045 proxy contracts) to gain a deeper understanding of the reasons for using the proxy pattern in DApp design. The minimum, median, maximum, and standard deviation of the number of proxies per DApp are 1, 2, 1,740, and 40.9, respectively. For each DApp, we clustered its proxy contracts based on its bytecode, as contracts with similar bytecode possess similar functionalities. We then selected a random proxy contract from each cluster. 

The first and third authors of this paper collaborated to study the source code of these proxies. Specifically, we employed a hybrid coding approach to discern the primary rationale behind using the proxy. As per source ERC-897 standard, a proxy contract can either be an upgradeability proxy or a forwarder proxy. The former is primarily utilized to upgrade DApp contracts, while the latter can serve various purposes such as cost-effective cloning of a contract that is expensive to deploy, logging, access control, routing, and more. Upon analyzing the proxy's source code, if it is used for upgradeability, we label it accordingly. On the other hand, if it is a forwarder proxy, we delve deeper into its source code to pinpoint a more specific reason. During this analysis, if we cannot match the reason with one from our existing codebook of reasons, we formulate a new one and subsequently incorporate it into our codebook.

\textbf{Findings}. \textit{Out of 50 studied DApps, 43 (86\%) used proxy contracts for upgradeability purposes, whereas 9 (14\%) with proxies that act as forwarders. Regarding the latter category, we found 6 systems that used proxy contracts towards reducing the deployment cost. The bold example is Uniswap-V1 for which we identified 1,740 proxy contracts. Uniswap-V1 uses the minimal proxy pattern to deploy a minimalist proxy per user. All deployed proxy contracts delegate to a singular logic contract address (i.e., Vyper\textunderscore contract), yet since every proxy executes the logic code in its own context, it allows reuse of the same logic code without the need to deploy the expensive logic per use. Furthermore, we found one DApp (i.e., Swerve Finance) that uses two forwarder proxy contracts to extend the functionality of the corresponding logic contracts. These specific forwarders add getter interfaces through which one can read the logic’s state variables. Finally, we found two other DApps (i.e., Element and Ether Legend) whose forwarder proxy contracts act as a hub/router, receiving requests and subsequently routing them to an appropriate logic contract.}

\section{Comparing on-Chain vs. off-Chain Deployment on gas costs and complexity}\label{subsec:gas-cost-complexity-of-dep-style}
In Observation \ref{obs:rq2-dep-style-and-contet-size}, we discovered that larger usage contexts typically favor on-chain deployment, while smaller ones lean towards off-chain methods. In this post-hoc analysis, our objective is to compare the average deployment gas cost across usage contexts for both deployment styles. We collected gas usage data for each proxy contract from the ``trace'' table in the Ethereum BigQuery dataset. We categorized the usage contexts of each deployment style based on their size (N) into either singleton (N=1) or multi-proxy (N>1) contexts. We then computed the average deployment gas cost for each context. For an on-chain context, this is the sum of the gas used for deploying its proxy contracts and the entire chain of factory contracts utilized in its creational pattern, divided by the number of deployed contracts. For an off-chain context, it is the sum of the gas used for deploying its proxy contracts divided by the number of deployed proxy instances. Additionally, we assessed the complexity of the off-chain and on-chain usage contexts by analyzing their bytecode length. We omitted factories from on-chain contexts in this analysis as we aimed to focus solely on the complexity difference in proxy contracts deployed through each style.

\begin{figure}[t]
\begin{minipage}[t]{0.49\columnwidth}
 \includegraphics[width=\linewidth]{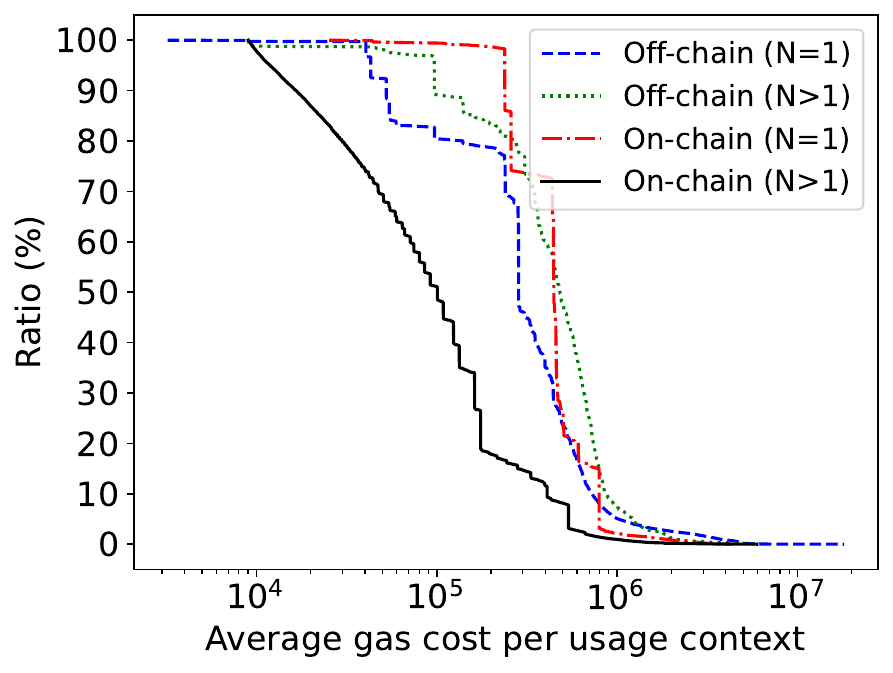}
 \caption{CCDFs of the average deployment gas cost per usage context of the top-5 on-chain and off-chain patterns.}
\label{fig:ccdf-avg-gas-cost-per-top-pattern}
\end{minipage}\hfill 
\begin{minipage}[t]{0.49\columnwidth}
 \includegraphics[width=\linewidth]{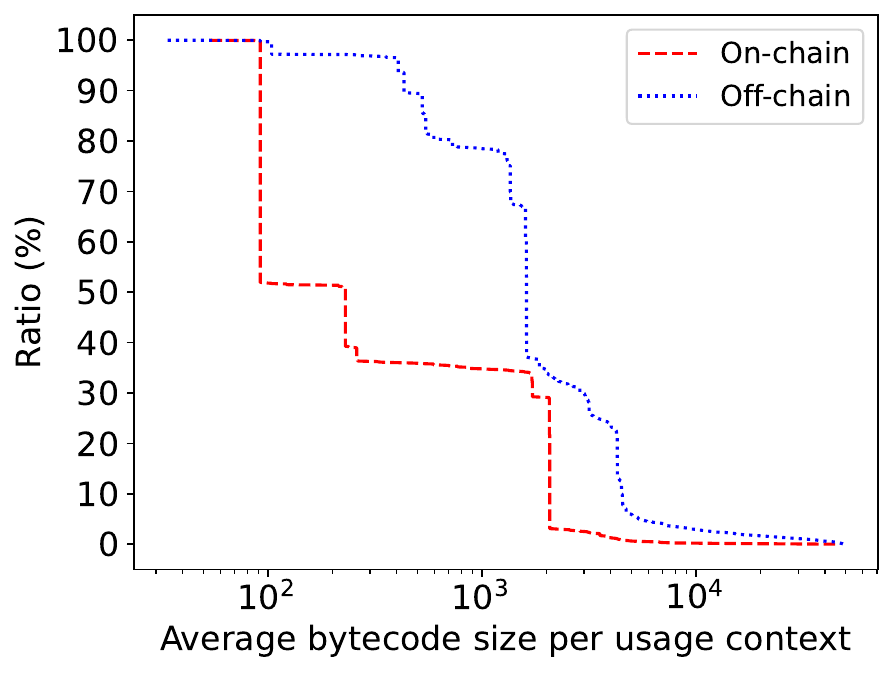}
 \caption{CCDFs of the average bytecode length per usage context of the on-chain and off-chain deployment style.}
\label{fig:ccdf-avg-bytecode-length-per-dep-style}
\end{minipage}
\end{figure}

\textbf{Findings.} \textit{Figure \ref{fig:ccdf-avg-gas-cost-per-top-pattern} presents our findings. We observed that on-chain (N=1) contexts (red line) typically consumed more gas during deployment than their off-chain (N=1) counterparts (blue line). A one-tailed Mann–Whitney U test indicates that this difference is statistically significant (P-value < 0.05) but with a small effect size (0.27). We hypothesize that the higher cost is attributable to the additional overhead of deploying factory contracts essential for on-chain singleton contexts. Interestingly, upon excluding the gas cost of these factory contracts from our calculations for on-chain (N=1) contexts, the average gas consumption dropped significantly below that of off-chain (N=1). This highlights the considerable overhead of factories when deploying a single proxy.}

\textit{Conversely, on-chain (N>1) contexts (black line) consistently consumed less gas than off-chain (N>1) contexts (green line), with a notable effect size of 0.72. We surmise that as the usage context size grows, the one-time gas expense for deploying factories remains more cost-effective for on-chain (N>1) contexts. Indeed, even after removing the gas cost for factories, on-chain (N>1) contexts still consumed significantly less gas than their counterparts. It is essential to note that while we are discussing the average deployment gas cost per usage context, the total deployment gas expense for on-chain contexts is usually higher than off-chain due to the broader scope of proxy deployment—exhibiting medians of 1,312,466 and 283,871, respectively. However, other elements, like initialization complexity and storage costs, might also affect deployment gas costs. Further investigation is needed to definitively assess the influence of these factors.}

\textit{Figure \ref{fig:ccdf-avg-bytecode-length-per-dep-style} reveals that off-chain usage contexts typically have a significantly larger bytecode size than on-chain contexts, with a substantial effect size of 0.52. This indicates that they offer more extensive functionality compared to on-chain contexts. We identified this pattern regardless of context size (i.e., whether N=1 or N>1). We speculate that since off-chain contexts are predominantly singleton (96\%) or smaller in scale (Observation \ref{obs:rq2-dep-style-and-contet-size}) compared to on-chain contexts, developers might prioritize adding functionalities over optimizing their proxy code. To delve deeper, we conducted a Spearman correlation test to examine the relationship between the size of the usage context (measured in terms of the number of proxy contracts) and its average bytecode size. The coefficient was -0.55, indicating a moderate inverse relationship between the two metrics. This supports our speculation that as the number of proxy contracts in a usage context increases, developers might opt for more gas-efficient proxy contracts (e.g., using minimal proxy contracts) to curtail deployment expenses. However, additional research is essential to confirm this causation.}

\section{Reducing deployment gas costs: A case study on ERC-1157 minimal proxy contracts}\label{apx:reducing-dep-cost-with-minimal proxy-contracts}
\smallskip \noindent \textbf{Approach}. To illustrate how cost-effective the proxy pattern can be in reducing deployment costs, we analyze the case of the Uniswap-V1 DApp (Appendix \ref{subsec:application-of-proxy-pattern-in-dapps}). We consider two scenarios: actual and hypothetical. In the first scenario (actual), the main logic contract called Vyper\textunderscore contract is deployed just once. Subsequently, 1,740 minimal proxy contracts are deployed, each of which delegates its operations to the Vyper\textunderscore contract. In the second scenario (hypothetical), the Vyper\textunderscore contract is directly and independently deployed 1,740 times. We recall that such an operation involves creating an entirely new and separate instance of the contract each time, encompassing both its logic and its state. 
    
We then computed the total deployment fee in Ether (ETH) for both scenarios. The transaction fee for deploying a contract is calculated by multiplying the number of used gas units with the gas price (in ETH) at the time of deployment. The gas price could vary based on several factors such as network congestion, miners' policy and network upgrades. Hence, for the first scenario, we scraped the deployment transaction fee and the corresponding gas price from each of the 1,740 minimal proxy contracts web pages on Etherscan. We then summed up the deployment transaction fees of all minimal proxy contracts and the deployment fee (0.04 ETH) of the singular Vyper\textunderscore contract to determine the total deployment fee.

Subsequently, we collected the number of gas units that were needed to deploy a single Vyper\textunderscore contract. Knowing the gas prices at the time of deploying the minimal proxy contracts from the first scenario, we multiplied them by the number of gas units required to deploy the Vyper\textunderscore contract. We then summed up all these calculations to derive the total deployment fee in the second scenario.

\textbf{Findings.} \textit{The total deployment gas cost for the first scenario (actual) is 2.07 ETH, which is approximately 14 times lower than that of the second scenario (28.46 ETH). Thus, our findings show that the careful use of proxy contracts can substantially reduce deployment gas costs at scale.
}

\section{Examples of largest DApps for the top-5 on-chain creational patterns}\label{apx:discussion-of-largest-dapps-for-the-top-5-patterns}
\smallskip \noindent \textbf{Approach}. The substantially large number of proxies (99.3\%) that are systematically deployed using on-chain factories motivated us to examine some examples of large decentralized systems that employ such patterns. Thus, we attempted to find the largest DApps associated with the top-5 on-chain creational patterns (i.e., the second to sixth rows of Table 2). For each of the top-5 patterns, we reused a similar method to the usage context viewpoint analysis in Section \ref{sec:rq1}, and then picked the largest usage context (a.k.a, the connected component) in terms of the number of proxy contracts. After which, we used Etherscan to manually identify the underlying system of the largest context. More specifically, we inspected the source/bytecode of the proxy contracts, logic contracts, and the contracts involved in the sequence through which the proxy is created. In particular, large-scale systems often utilize the name of the decentralized system as a pseudonym for the deployer address, which facilitates our identification of them in certain cases. Following, we discuss each pattern in the context of its corresponding system.

\begin{enumerate}[itemsep = 3pt, topsep = 0pt]
    \item \textbf{EOA > FA > P}. We identified 6,618,012 proxy contracts for this pattern. The largest usage context we found for this pattern, is Opensea\footnote{\url{https://opensea.io}} with 943,022 proxies (14.2\%). Opensea is the most prominent digital marketplace for crypto collectibles, especially non-fungible tokens (NFTs). Opensea uses a simple factory contract (WyvernProxyRegistry) to deploy one proxy per user, allowing them to have control over their assets. Each item that is traded on Opensea is owned by the proxy smart contract of a user. The corresponding sequence of contract names is as follows. Concerning the proxy (factory) contracts, the proxy contract name and its logic name are presented as a pair (e.g., [proxy contract name, logic contract name]). 
    
    \smallskip
    \textit{EOA > FA [WyvernProxyRegistry\footnote{\url{https://etherscan.io/address/0xa5409ec958c83c3f309868babaca7c86dcb077c1}}] > P [OwnableDelegateProxy\footnote{\url{https://etherscan.io/address/0xf69f020053685724feda8e617486442e08995607}}, AuthenticatedProxy\footnote{\url{https://etherscan.io/address/0xf9e266af4bca5890e2781812cc6a6e89495a79f2}}]}
    
    \item \textbf{EOA > PF > P}.  We identified 379,293 proxy contracts for this pattern. The largest usage context we found for this pattern, is Kraken\footnote{\url{https://www.kraken.com}} with 202,151 proxies (53.3\%).  Kraken is a decentralized exchange that provides a trading platform for buying, selling, and trading crypto assets. Kraken uses a proxy factory contract whose logic can be updated at runtime. Using this proxy factory, Kraken deploys a \textit{minimalistic proxy} for each user (see EIP-1167 in Section~\ref{sec:background}). This minimalistic proxy merely forwards calls to a logic contract. Interestingly, the logic contract itself is a proxy contract that conducts several input validation checks before forwarding the calls to the final logic contract. Therefore, proxies here act as gates that validate the transaction input data. Our findings also show that there exits complex systems that use several layers of proxies stacked on top of each other. The corresponding sequence of contract names is as follows. 

    \smallskip
    \textit{EOA > PF [Non Verified\footnote{\url{https://etherscan.io/address/0xf9e266af4bca5890e2781812cc6a6e89495a79f2}}, Not Verified\footnote{\url{https://etherscan.io/address/0x2febe3d60bb2bbdab135c740617241c2eb949635}}] > P [Not Verified\footnote{\url{https://etherscan.io/address/0x439039fbbbe5fd1de44fcacca694ac0da197a624}}, Not Verified\footnote{\url{https://etherscan.io/address/0x83b76b11257c4ece35370b6152f1946d49479e89}}]}

    \item \textbf{EOA > FA > FA > P}. We identified 123,349 proxy contracts for this pattern. The largest usage context we found for this pattern is the GnosisSafe Wallet DApp\footnote{https://safe.global/}, which contains 51,746 (42\%) proxy contracts. The corresponding sequence of contract names is as follows.

    \smallskip
    \textit{EOA > FA [Generic Factory\footnote{\url{https://etherscan.io/address/0x4e59b44847b379578588920ca78fbf26c0b4956c}}] > FA [GnosisSafeProxyFactory\footnote{\url{https://etherscan.io/address/0xa6b71e26c5e0845f74c812102ca7114b6a896ab2}}] > P [GnosisSafeProxy\footnote{\url{https://etherscan.io/address/0xec55b27ffce0ed7c1de6568d98ad9a3b6219b6a7}}, GnosisSafe\footnote{\url{https://etherscan.io/address/0xd9db270c1b5e3bd161e8c8503c55ceabee709552}}]}

    \smallskip
    The instances of GnosisSafeProxy contracts are deployed by the GnosisSafeProxyFactory, which instantiates one instance per user request. We found that developers used another FA to deploy the GnosisSafeProxyFactory itself. We further analyzed why developers used another FA to deploy GnosisSafeProxyFactory itself. We found that the factory used for deploying GnosisSafeProxyFactory is indeed a generic factory and with minimalistic code. It only receives a bytecode and then deploys it using the \texttt{create2} statement. Listing \ref{listing:general-purpose-factory} shows the decompiled bytecode of this factory. Therefore, the chain becomes longer because developers preferred to use an on-chain generic factory to deploy the specialized GnosisSafeProxyFactory instead of off-chain scripts. There can be several reasons for using such a generic on-chain factory, including the determinism offered by the \texttt{create2} opcode, ensuring consistency across platforms and Ethereum forks, and simplifying deployment logic for contract reusability.
\begin{figure}[t]
\begin{lstlisting}[frame=single, style=sidebyside, language=Solidity, caption={The decompiled bytecode of the general purpose factory used by GnosisSafe.},label={listing:general-purpose-factory}]
def _fallback() payable: # default function
  create2 contract with callvalue wei
                  salt: call.func_hash
                  code: call.data[32 len calldata.size - 32]
  require create2.new_address
  return addr(create2.new_address)
\end{lstlisting}
\end{figure}
    \item \textbf{EOA > FA > FA > FA > FA > P}. We identified 35,764 proxy contracts for this pattern, primarily associated with the Dharma Wallet DApp\footnote{https://www.dharma.io/} with 18,397 proxy contracts. The corresponding sequence of contract names is as follows: 

    \smallskip
    \textit{EOA > FA [Generic Factory\footnote{\url{https://etherscan.io/address/0x7a0d94f55792c434d74a40883c6ed8545e406d12}}] > FA [ImmutableCreate2Factory\footnote{\url{https://etherscan.io/address/0xcfa3a7637547094ff06246817a35b8333c315196}}] > FA [ImmutableCreate2Factory\footnote{\url{https://etherscan.io/address/0x0000000000ffe8b47b3e2130213b802212439497}}] > FA [DharmaSmartWalletFactoryV1\footnote{\url{https://etherscan.io/address/0xfc00c80b0000007f73004edb00094cad80626d8d}}] > P [UpgradeBeaconProxyV1\footnote{\url{https://etherscan.io/address/0x0004c95f9ba50a1ea11544565b71fab5dc5658c0}}, DharmaSmartWalletImplementationV15\footnote{\url{https://etherscan.io/address/0x4d90ca10f218ebd4509ca0f9816cf20fd9903c35}}]}
    \smallskip
    In this example, an EOA deployed a generic factory contract, followed by the deployment of the ImmutableCreate2Factory using the generic factory. ImmutableCreate2Factory offers secure contract deployment with a salt value and initialization code, preventing redeployment, collisions, and front-running. Another instance of ImmutableCreate2Factory is deployed through the former ImmutableCreate2Factory, and eventually, the DharmaSmartWalletFactoryV1 factory is deployed via the latter ImmutableCreate2Factory, which deploys clones of UpgradeBeaconProxyV1 proxy contracts. We further analyzed who initiated each of the four factory contracts' creation transactions:
    \begin{itemize}[label=\textbullet, itemsep = 3pt, topsep = 0pt]
        \item EOA initiated the Generic Factory (FA) creation transaction.
        \item EOA2 initiated the first ImmutableCreate2Factory (FA) creation transaction.
        \item EOA2 also initiated the second instance of the ImmutableCreate2Factory creation transaction.
        \item Dharma initiated the DharmaSmartWalletFactoryV1 creation transaction.
    \end{itemize}
    Given this analysis, it is evident that three different EOAs initiated the creation transactions of the intermediary factory contracts. Assuming that each EOA represents an independent developer, this observation mainly highlights the practice of developers/practitioners reusing factory contracts deployed by others for their own purposes, which can lead to a higher pattern length.

    \item \textbf{EOA > PF > PF > P}. We identified 1,968 proxy contracts, primarily associated with the PlotX DApp\footnote{https://plotx.io/} with 720 (36.5\%) proxy contracts. The corresponding sequence of contract names is as follows. 

    \smallskip
    \textit{EOA > PF [OwnedUpgradeabilityProxy\footnote{\url{https://etherscan.io/address/0x03c41c5aff6d541ef7d4c51c8b2e32a5d4427275}}, Master\footnote{\url{https://etherscan.io/address/0x4eb82f80858e6974307458b057b875bf3f23c4a0}}] > PF [OwnedUpgradeabilityProxy\footnote{\url{https://etherscan.io/address/0xe210330d6768030e816d223836335079c7a0c851}}, MarketRegistryNew\footnote{\url{https://etherscan.io/address/0x495d3a0530367ed4331833eae74b32d4848401f0}}] > P [OwnedUpgradeabilityProxy\footnote{\url{https://etherscan.io/address/0x4c7861e96dc5a6e5acb31833ea5722e800a7bd24}}, Market\footnote{\url{https://etherscan.io/address/0x25cf9d73b711bff4d3445a0f7f2e63ade5133e67}}]}
    
    \smallskip
    In this instance, the EOA deploys the first OwnedUpgradeabilityProxy ``PF'' contract, enabling upgradeability for the Master logic contract, which serves as a central registry and controller for other contracts within the PlotX DApp. The second OwnedUpgradeabilityProxy ``PF'' contract enables upgradeability for the MarketRegistryNew logic contract, primarily dealing with market creation and reward distribution. The MarketRegistryNew deploys Market instances by instantiating another OwnedUpgradeabilityProxy proxy contract that delegates to a Market logic contract. Our analysis revealed that all of these intermediary ``PF'' contracts were initiated by the PlotX organization, indicating the use of multiple layers of upgradeability proxy contracts for managing and upgrading their DApps' contracts at different levels.

\end{enumerate}

\section{Proxy classification process}
\label{apx:proxy-classification-process}

\begin{figure}[!htbp]
    \centering
    \includegraphics[width=1\columnwidth]{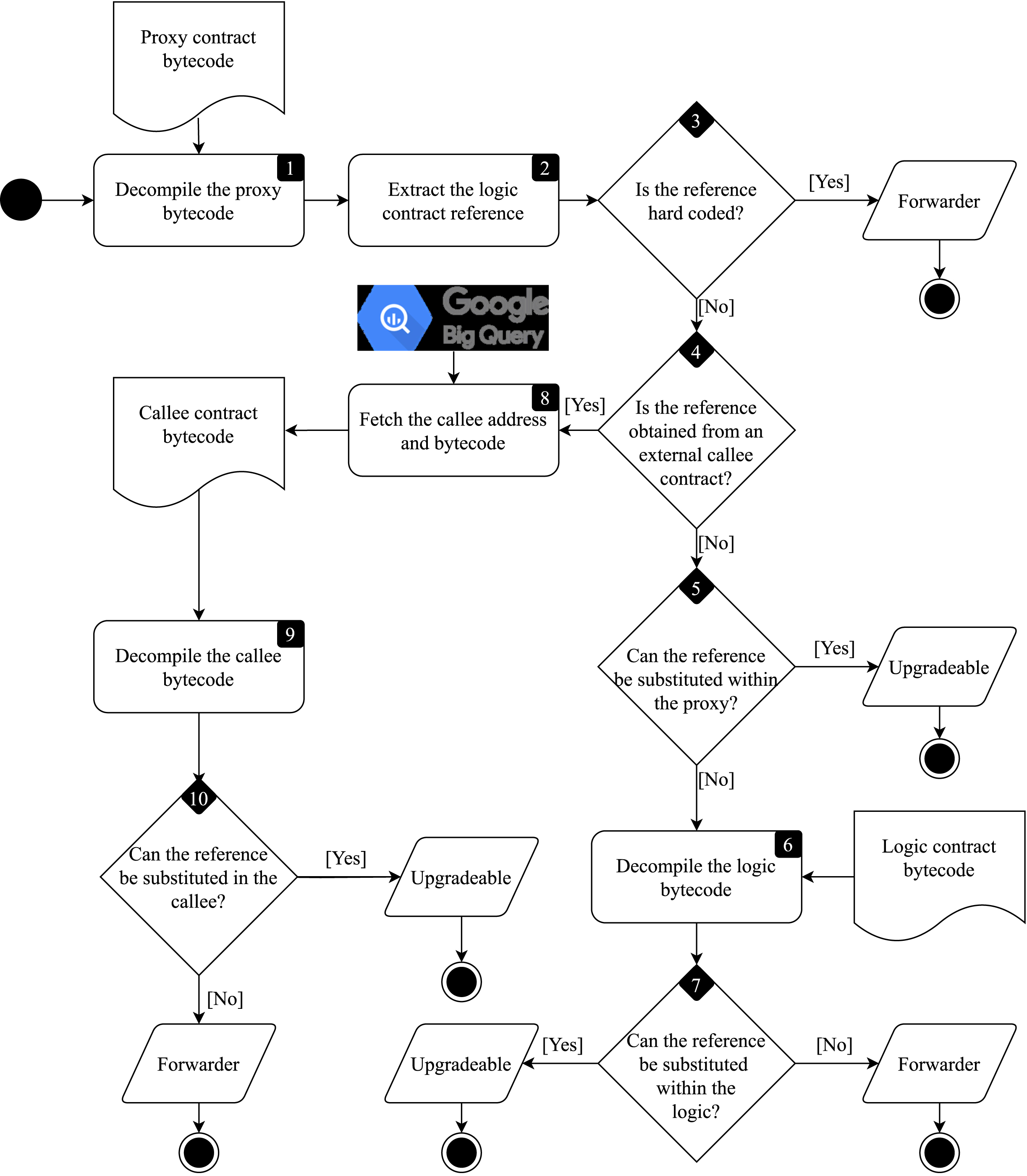}
    \caption{The process of classifying a proxy contract into upgradeability and forwarder classes.}
   \label{fig:proxy-classification}
\end{figure}

We followed an approach proposed by \citet{Salehi22} to classify proxy contracts into either forwarder or upgradeability classes. We studied their approach, extracted the workflow, and adopted it in our qualitative study protocol in Section \ref{sec:rq3}. Figure \ref{fig:proxy-classification} depicts the process used for manually classifying proxy contracts into upgradeability and forwarder types. The input of the process is a proxy contract and its logic contracts. Below, we summarize the ten steps involved along with examples. Note that, due to limited space, the examples are excerpted from real-world contacts. We use the ``[...]'' to summarize the source code. 

\begin{enumerate}[itemsep = 3pt, topsep = 0pt]
    \item The approach relies on the contract's decompiled bytecode. Therefore, we decompile the proxy contract's bytecode using the Panoramix tool.
    
    \item We extract the logic contract's reference variable from the function where the proxy functionality is implemented. Listing \ref{listing:upgradeability-proxy} shows an example of a proxy. The proxy functionality is implemented between 8 to 18. In this example, the logic contract's reference variable can be found at like 11, \texttt{stor3608.field\textunderscore 0}.
    
    \item After identifying the reference of logic contract, we check whether it is hard-coded. In Listing \ref{listing:upgradeability-proxy}, the reference is not hard-coded because its definition can be found in the \texttt{storage} function (line 3). However, Listing \ref{listing:forwarder-proxy-1} provides an example where the logic contract reference is hard-coded in line 2. In the former case, we proceed to step 4, while in the latter case, we label the proxy as a forwarder.
    
    \item We check where the logic reference variable is obtained through performing a call to an external callee contract. If so, we proceed to step 8. Otherwise, We record the storage slot of the reference variable and then proceed to step 5. In Listing \ref{listing:upgradeability-proxy}, the reference and its reserved storage slot (\texttt{0x36[...]bc}), is defined in the proxy contract storage in line 3. On the other hand, Listing \ref{listing:beacon-proxy-1} shows an example where the proxy makes a call to a callee to inquire about the logic reference in line 7. The proxy then reads the logic address from the returned value of the external call and then delegates the call in line 11.
    
    \item If the reference variable is found in the proxy contract, we then check whether there is a function that allows developers to substitute the old reference with an updated one. To do this, we check all variable assignments and filter those whose left operand includes the reference variable. Next, we evaluate each assignment individually. In particular, we examine the function where the assignment occurs to see if it receives an argument that is used as the right operand in the assignment. In the positive case, it means that the function allows updating the logic reference; thus, we label the proxy as an upgradeability proxy. Otherwise, we proceed to step 6. Listing \ref{listing:upgradeability-proxy} is an example where there are two functions that enable substituting the logic reference. The first one is \texttt{updateTo} function (line 25 to line 32) and the logic reference is updated in line 32 by a function's argument. The second function (\texttt{updateToAndCall}) is defined between lines 35 to 42 where the logic reference is updated in line 41.
    
    \item It is possible that the function that enables substituting the logic reference gets implemented in the logic contract rather than the proxy. In this scenario, the proxy contract delegates a call to the update function implemented within the logic contract. Since the execution happens in the context of the proxy with \textit{delegatecall} (Section \ref{sec:background}), when the logic contract updates the logic reference, it indeed substitutes the logic reference stored in the proxy contract storage, and the next delegatecall will use the new logic contract address. Thus, if we cannot find it within the proxy, we then search for it in the logic contract. We decompile the logic contract's bytecode.
    
    \item Having known the storage slot of the reference variable (step 3), we then check if the logic contract has a variable in its storage with a similar storage slot. If so, we extract the variable name and repeat the fifth step to assess if there is a function within the logic contract that updates the reference variable. In the positive case, we label the proxy as an upgradeability proxy; otherwise, we categorize it as a forwarder. Listing \ref{listing:uups-proxy-1} shows an example of a proxy and its logic contract where the latter implements the update function.
    
    \item If the logic reference comes from an external callee contract, then the update function can be implemented in the callee. To check this, we first need to find the callee's address. Specifically, we use the BigQuery dataset to analyze the proxy's traces and fetch the trace related to the external call, i.e., the last trace that occurs just before the delegatecall trace. We then extract the callee from the fetched trace. Additionally, we extract the first four bytes from the trace's calldata, i.e., the function selector of a function that returns the logic reference to the proxy. We refer to this function as the ``getter'' function. We also extract the callee's bytecode in this phase.
    
    \item We then decompile the callee's bytecode.
    
    \item Having known the getter function selector, we then look for a function with a similar selector within the callee contract. If we find a match, we check the function's return statement and record the returned variable (i.e., the reference variable). Once we detect the reference variable, we repeat the fifth to figure out if there is a function within the callee contract that allows substituting the reference variable. If there is such a function, we label it as an upgradeability proxy; otherwise, we label it as a forwarder proxy. Listing \ref{listing:beacon-proxy-1} shows an example of a proxy and an external callee contract. The proxy calls the \texttt{implementation()} function from the external callee in line 7. The \texttt{implementation()} getter function is implemented between lines 23 and 24 in the callee contract, where it returns the \texttt{implementationAddress} reference variable in line 24. Having applied the fifth step, we can observe that the callee contract implements an \texttt{upgradeTo} function that allows updating the \texttt{implementationAddress} with a new implementation address in line 32. Therefore, we label the proxy as an upgradeability proxy.
    
\end{enumerate}

\noindent
\begin{figure}[ht]
\begin{lstlisting}[frame=single, style=sidebyside, language=Solidity, caption={An example of upgradeability proxy.},label={listing:upgradeability-proxy}]
def storage:
  stor3608 is uint128 at storage 0x36[...]bc offset 160
  stor3608 is addr at storage 0x36[...]bc
  stor3608 is uint256 at storage 0x36[...]bbc
  [...]

# the function that implement the proxy functionality
def _fallback() payable: # default function
  if caller == addr(storB531.field_0):
      [...]
  delegate uint256(stor3608.field_0) with: # the reference variable
     funct call.data[0 len 4]
       gas gas_remaining wei
       args call.data[4 len calldata.size - 4]
  
  if not delegate.return_code:
      revert with ext_call.return_data[0 len return_data.size]
  return ext_call.return_data[0 len return_data.size]

# return the logic contract address
def implementation(): # not payable
  [...]

# substitute the logic reference with a new reference logic contract
def upgradeTo(address _implementation): # not payable
  require calldata.size - 4 >= 32
  if addr(storB531.field_0) != caller:
      [...]
  if ext_code.size(_implementation) <= 0:
      [...]
  addr(stor3608.field_0) = _implementation # substitute the logic reference
  [...]

# substitute the logic reference with a new reference logic contract
 and then make a call.
def upgradeToAndCall(address _implementation, bytes _data) payable: 
  require calldata.size - 4 >= 64
  require _data <= 4294967296
  require _data + 36 <= calldata.size
  [...]
  addr(stor3608.field_0) = _implementation # substitute the logic reference
  [...]

# return the admin address
def admin(): # not payable
  [...]

# change proxy admin to a new address
def changeAdmin(address _admin): # not payable
  [...]

\end{lstlisting}
\end{figure}

\noindent
\begin{figure}[t]
\begin{lstlisting}[frame=single, style=sidebyside, language=Solidity, caption={An example of a forwarder proxy where the logic contract reference is hard coded at line 3.},label={listing:forwarder-proxy-1}]
def _fallback() payable: # default function
  delegate 0x2f5e324ec0e2fd9925165c66e0daade39837adb5 with:
     funct call.data[return_data.size len 4]
       gas gas_remaining wei
       args call.data[return_data.size + 4 len calldata.size - 4]
  if not delegate.return_code:
      revert with ext_call.return_data[return_data.size len return_data.size]
  return ext_call.return_data[return_data.size len return_data.size]
\end{lstlisting}
\end{figure}

\noindent
\begin{figure}[t]
\begin{lstlisting}[frame=single, style=sidebyside, language=Solidity, caption={An example of a proxy where the logic contract reference is returned by an external callee.},label={listing:beacon-proxy-1}]
======================== Proxy contract =======================
def storage:
  storA3F0 is addr at storage 0xa3[...]50

def _fallback() payable: # default function
  require ext_code.size(storA3F0)
  static call storA3F0.implementation() with: # external call
          gas gas_remaining wei
  [...]
  # read the logic reference from the returned value
  delegate ext_call.return_data[0] with: 
     funct call.data[0 len 4]
       gas gas_remaining wei
      args call.data[4 len calldata.size - 4]
  [...]

=================== External callee contract ==================
def storage:
  owner is addr at storage 0
  implementationAddress is addr at storage 1

# the reference variable getter function
def implementation() payable: 
  return implementationAddress # reference variable

# substitute the logic reference with a new reference logic contract
def upgradeTo(address _implementation) payable: 
  require calldata.size - 4 >= 32
  if owner != caller:
      revert with 0, 'Ownable: caller is not the owner'
  [...]
  implementationAddress = _implementation
  [...]

# return the proxy admin
def owner() payable: 
  [...]
 
# renounce the ownership
def renounceOwnership() payable: 
  [...]

# transfer the ownership to another address
def transferOwnership(address _newOwner) payable: 
  [...]



\end{lstlisting}
\end{figure}

\noindent
\begin{figure}[t]
\begin{lstlisting}[frame=single, style=sidebyside, style=sidebyside, language=Solidity, caption={An example of a proxy and it logic contract. The logic implements the update function rather than the proxy.},label={listing:uups-proxy-1}]
======================== Proxy contract =======================
def storage:
  # The storage slot of the reference variable is 0x36[...]bc
  stor3608 is addr at storage 0x36[...]bc

def _fallback() payable: # default function
  delegate stor3608 with:
     funct call.data[0 len 4]
       gas gas_remaining wei
      args call.data[4 len calldata.size - 4]
  [...]

======================== Logic contract =======================
def storage:
  [...]
  # exactly similar storage slot.
  stor3608 is addr at storage 0x36[...]bc
  [...]
 
# Update the logic contract reference to a new reference
# logic contract. Having called by the proxy, this changes
# the proxy's storage, leading to the substitution of the
# logic reference variable in the proxy storage.
def upgradeTo(address _implementation): # not payable
  require calldata.size - 4 >= 32
  [...]
  addr(stor3608) = _implementation # substitute the logic reference
  if not stor4910:
      [...]
      addr(stor3608) = _implementation # substitute the logic reference
      log Upgraded(address nextVersion=_implementation)
\end{lstlisting}
\end{figure}

    \end{appendices}
\end{document}